\documentclass[prb,aps,showpacs,citeautoscript, superscriptaddress,amsmath,amssymb,floatfix,twocolumn,dvipsnames,longbibliography,nobibnotes]{revtex4-1}
\usepackage{graphicx}
\usepackage{xcolor}
\usepackage{ytableau}
\usepackage{amsfonts}
\usepackage{amsmath}
\usepackage{times}
\usepackage{amssymb}
\usepackage{changes}
\usepackage{blkarray}
\usepackage{relsize}
\usepackage{multirow}
\usepackage{makecell}
\usepackage{amsthm}

\usepackage{comment}
\usepackage{tcolorbox}

\usepackage{algpseudocode}

\ytableausetup{smalltableaux} 

\setcitestyle{super}

\DeclareRobustCommand{\SU}[1]{\ensuremath{\mathrm{SU}(#1)}}

\newcommand{\shsc}[3]{\mathrel{\raisebox{#1}{\protect\scalebox{#2}{#3}}}}

\usepackage{color}
\definecolor{lightpink}{RGB}{255, 187, 218}
\definecolor{lightblue}{RGB}{130, 215, 255}
\definecolor{lightgrey}{RGB}{220, 220, 220}
\definecolor{lightred}{RGB}{255, 155, 155}
\definecolor{lightteal}{RGB}{143, 255, 243}
\definecolor{lightindigo}{RGB}{191, 181, 255}
\definecolor{lightgreen}{RGB}{185, 255, 179}
\definecolor{lightpurple}{RGB}{237, 179, 255}

\definecolor{c1}{RGB}{144,12,63}
\definecolor{c2}{RGB}{199,0,57}
\definecolor{c3}{RGB}{255,87,51}
\definecolor{c4}{RGB}{255,195,0}
\definecolor{c5}{RGB}{218,247,166}

\usepackage[
colorlinks=true,
bookmarks=false,
citecolor=blue,
linkcolor=blue,
urlcolor=blue
]{hyperref}

\begin{document}

\title{Spectral Mapping Theorem between \SU{N} and \SU{N_s/N} Heisenberg Models and Exact Diagonalization of \SU{4} and \SU{7} Long-Range Ordered Phases on the Triangular Lattice}

\date{\today} 
\author{Samuel Gozel}
\thanks{Independent researcher}
\email[Contact author: ]{samuel.gozel@alumni.epfl.ch}
\affiliation{Switzerland}

\author{Pierre Nataf}
\affiliation{Laboratoire de Physique et Mod\'elisation des Milieux Condens\'es, Universit\'e Grenoble Alpes and CNRS, 25 avenue des Martyrs, 38042 Grenoble, France}

\begin{abstract}
We first show that, for the \SU{N} Heisenberg model with isotropic two-site interactions on an arbitrary lattice of $N_s$ sites, the full spectrum in its singlet sector (when $N$ divides $N_s$) is the negative of the full spectrum in its singlet sector of the \SU{N_s/N} Heisenberg model defined on the same lattice.
This theorem, valid for models with the fundamental irreducible representation (irrep) at each lattice site, follows naturally from a transposition operation in the basis of Standard Young Tableaux (SYTs).
Second, this basis -- previously recognized as particularly convenient for exact diagonalization (ED) of \SU{N} Heisenberg models -- admits an efficient, memory-saving implementation, thereby enabling large-scale diagonalizations for unprecedented system sizes $N_s$ and numbers of colors $N>3$.
We apply these ideas to an ED study of the \SU{4} and \SU{7} Heisenberg models on the triangular lattice (TL), extending the calculations up to $N_s=28$ sites.
In particular, starting from the \SU{4} case, whose nearest-neighbor ground state remains debated, we confirm that next-nearest-neighbor couplings stabilize four-sublattice long-range order. We further find that increasing the number of colors from $N=4$ to $N=7$ provides another route toward a N\'eel-like ground state, even without such additional couplings.

\end{abstract}

\maketitle

 \section{Introduction}

The model that Werner Heisenberg introduced to study ferromagnetism in 1928 is arguably the most important model in quantum magnetism~\cite{Heisenberg_1928}.
It has not only proved useful for describing quantum spin systems, but has also motivated a plethora of analytical and numerical developments
that go far beyond the study of interacting spin-$1/2$ degrees of freedom on a lattice. One example is the Bethe ansatz: initially devised to solve exactly the spin-$1/2$ Heisenberg model (HM) on a chain~\cite{Bethe_1931}, it was later generalized to other integrable models, giving rise to an extensive branch of mathematical physics~\cite{Essler2005,Caux_2011}.

In all these developments, the symmetries of the model have remained central: they provide physically relevant quantum numbers for the eigenstates~\cite{chen,Tinkham2003,Dresselhaus2008} and reduce the complexity of solving the Schr\"odinger equation by block-diagonalizing the Hamiltonian into independent symmetry sectors~\cite{Dagotto1994,Lin1993,laflorencie_2008_simulations,
frustrated_book,
weisse_2013_divide,
wietek_2018_sublattice}.

In this respect, the mathematical connection between the HM and the permutation group has always been particularly fruitful.

As early as 1929, Paul Dirac noted that the isotropic exchange interaction between two spin-$1/2$ degrees of freedom at sites $i$ and $j$, $\hat{S}_i \cdot \hat{S}_j$,
can be written
%with 
as a
{\it permutation} operator, $\hat{S}_i \cdot \hat{S}_j=(1/2) P_{i,j}-1/4$, where $P_{i,j}$ exchanges the spin states (spin up or down) at sites $i$ and $j$~\cite{Dirac_1929}.
Later, Powers reinterpreted the isotropic HM as a random walk on the permutation group~\cite{Powers1976}, while Sutherland used permutations between $N$-component systems on nearest-neighbor sites of a chain to study an integrable model that he solved via the Bethe ansatz~\cite{Sutherland1975}.
Sutherland and other experts on integrable models ~\cite{Kulish_1981} also noted the \SU{N} invariance of the interactions in these permutation-based models of multicomponent systems, leading to what is now called the \SU{N} HM.

Later, using ideas from high-energy physics and motivated by the description of spin-$1/2$ fermions, Papanicolaou and Affleck developed large-$N$ approaches to \SU{N} HMs~\cite{Papanicolaou_1984,
Papanicolaou1988,
Affleck_1985_largeN,
Affleck_1988_largeN,
Marston_1989_largeN}. 
At the beginning of the 2000s, it was recognized that \SU{N} HMs could be realized using ultracold atoms trapped in optical lattices~\cite{wu_exact_2003,
Wu_review_2006,
gorshkov_two_2010,
Cazalilla_2014,
Capponi_annals_2016,
Ibarra_Garcia_Padilla_2024}. 
As a result, what had long been regarded as a purely theoretical model became experimentally accessible. In these systems, $N$ is determined by the number of accessible nuclear-spin states and can reach values as large as $N=10$~\cite{takahashi2012,Pagano2014,Scazza2014,Zhang2014,Hofrichter_2016,Becker_2021,taie2020observation,Fallani_2022,Pasqualetti_2024,tarruell_2026,salomon_2026}.

Theoretical investigations of \SU{N} HMs subsequently flourished, driven by the goal of determining the conditions under which exotic phases might arise, thereby extending the physical phenomena observed in their \SU{2} counterparts. Depending on the lattice geometry and the number of colors $N$, numerous two-dimensional phases have been theoretically predicted at zero temperature. These include: N\'eel long-range color-ordered (LRO) states~\cite{Corboz_2011,Bauer_2012}, which are the \SU{N} equivalents of the well-known ($\pi, \pi$) (respectively $120^{\circ}$) N\'eel states existing on the square lattice~\cite{Anderson_1952,Manousakis_1991} (respectively TL ~\cite{Lecheminant_1994,Sorella_1999}) for $N=2$; 
plaquette phases~\cite{paramekanti_2007,
corbozsimplex2012,
Assad_2013,
natafrapidcomm2016,
Gauthe_2021,
Jin_2022}, which can be interpreted as generalizations of valence-bond states for spin-$1/2$ systems; and various types of spin-liquid phases~\cite{Wang_2009,
Hermele_2011,
Corboz_prx_2012,
Lai_2013,
Chen_Hazzard_2016,
natafPRL2016,
Boos_2020,
Keselman_2020,
Chen_2021}, which constitute higher-symmetry generalizations of the Anderson resonating valence-bond (RVB) states~\cite{Anderson_1973_Resonating}.

 These theoretical studies are hampered by the exponential growth of the full Hilbert-space dimension: in the simplest case of a fundamental \SU{N} irreducible representation (irrep) of dimension $N$ on each site~\cite{weyl,chen,nataf2014}, such a dimension scales as $N^{N_s}$, where $N$ denotes the number of colors and $N_s$ the number of lattice sites.
 Consequently, implementing symmetries, and in particular \SU{N} symmetry, became essential~\cite{Sierra_1996,mcculloch2002,weichselbaum2012,Bruognolo_2021}. In particular, working in the \SU{N} singlet sector, formed by many-body wave functions invariant under global \SU{N} rotations and containing the low-lying states in the antiferromagnetic case, has proved particularly fruitful. It yields a dimensional reduction of several orders of magnitude, which increases with $N$ at fixed $N_s$~\cite{nataf2014}.
 Unsurprisingly, such an implementation -- 
 %painful 
 cumbersome 
 but feasible with \SU{N} Clebsch-Gordan coefficients (CGCs)~\cite{Nagel1989CGRacah,Kota2004SUNRacah,alex_2011_anumerical,weichselbaum2020} -- proved particularly straightforward using the representation theory of the permutation algebra~\cite{rutherford}, which Alfred Young developed in the 1930s and 1940s~\cite{Young6}.
 
Such a connection between the \SU{N} HM and the permutation group, already pointed out case by case at the level of physical models, has its mathematical origin in Schur--Weyl duality~\cite{Rowe_2012,Bjornberg_2023,Polychronakos_2023}.
In this framework, permutations of the symmetric group $\mathcal{S}_{N_s}$ naturally act on the factors of the tensor product of local \SU{N} irreps. 
They decompose the full Hilbert space into multiple copies of independent sectors invariant under the Hamiltonian and characterized by a given \SU{N} symmetry~\cite{Schur1901,weyl,FultonHarris1991,cornwell,Hamermesh1962}.

The \SU{N}-invariant HM Hamiltonian is reinterpreted as an element of the algebra of $\mathcal{S}_{N_s}$, the \SU{N}-invariant sectors are labeled by Young diagrams~(YD), 
and when these YD are filled up with site (or particle) indices in the {\it standard} way, one obtains a very convenient {\it basis} of Standard Young Tableaux (SYTs), 
on which the matrix elements of the Hamiltonian and of the \SU{N}-invariant observables are easily computed through the rules of the orthogonal representation of $\mathcal{S}_{N_s}$~\cite{Young6,rutherford,nataf2014}. 
Although the basis of SYTs was already known in chemistry for the {\it electron} problem (i.e. for $N=2$)~\cite{Goddard_1967,Gallup_1973,Salmon_1974}, it has proved particularly useful and efficient when $N \geq 3$, 
as alternative CGC-based methods become computationally prohibitive~\cite{Bruognolo_2021,
QSpace_1,
Vanderstraeten_2019,
Fishman_2022}.
 
Originally used to compute the low-energy spectra of \SU{N} HMs via exact diagonalization~ (ED)~\cite{nataf2014,PRBNataf2016,Wan2017}, 
the basis of SYTs has already played a significant role in advancing our understanding of \SU{N} quantum magnetism. 
It has, for instance, been instrumental in the study of the \SU{3} chiral spin liquid on the TL~\cite{Boos_2020}, 
in establishing that the \SU{5} HM on the square lattice exhibits long-range N\'eel order~\cite{nataf2014}, 
and in demonstrating that the \SU{6} model on the honeycomb lattice realizes a plaquette phase~\cite{natafrapidcomm2016}.
 
Because it naturally accommodates (\SU{N}-invariant) bond disorder, it was also employed to examine the behavior of mid-spectrum eigenstates in \SU{3} disordered chains~\cite{Dabholkar_2024} and to study the spin-mixing dynamics of a one-dimensional strongly repulsive Fermi gas under harmonic confinement~\cite{Pecci_2022}.
It was additionally used to obtain the thermal properties of \SU{N} systems: the temperature dependence of spin correlations in the \SU{3} (respectively \SU{4}) HM on the TL or square lattice for $N_s=18$ (respectively $N_s=16$) sites~\cite{Romen_2020}, 
and the specific heat of the \SU{4} HM on the $N_s=24$-site honeycomb lattice~\cite{Yamada_2022}.
 
The purpose of the present article is twofold: first, to show that SYTs can also be used at a more {\it mathematical} level, 
as the key ingredient in the proof of a simple and general mapping theorem connecting the singlet spectra 
of the antiferromagnetic \SU{N} HM to those of the ferromagnetic \SU{N_s/N} HM on the same $N_s$-site lattice.
Second, we
%wish to give 
propose 
a {\it computationally} efficient way to store and manipulate the basis of SYTs.
As a main application of this latter development, and as a striking illustration of the mapping theorem for $N_s=28$, we study the \SU{4} and \SU{7} HM ground states on the TL.
The paper is organized as follows.
  
In Sec.~\ref{theorem}, after a brief review of the SYT-based procedure for the ED of the \SU{N} HM on finite-size clusters, we state and prove the mapping theorem.
Then, in Sec.~\ref{tech_aspec}, we describe an efficient strategy to index and store SYTs in their Yamanouchi representation~\cite{Yamanouchi1937}. Implementation details and performance benchmarks are reported in Appendix~\ref{app:computational_details}.
Finally, in Sec.~\ref{LRO}, we test our improved algorithm on the \SU{N} HM on the TL: first, in Sec.~\ref{LRO4}, we extend previous results on the four-sublattice long-range ordered state for $N=4$ with next-nearest-neighbor (NNN) couplings to larger clusters; then, in Sec.~\ref{LRO7}, we show that $N=7$ colors naturally favor LRO on the TL without NNN couplings.

\section{Mapping theorem}
\label{theorem}
\subsection{Review of the SYT-based exact diagonalization of Heisenberg \SU{N} models}
\paragraph{The model}

The \SU{N} HM Hamiltonian takes the form
\begin{align}
\label{hamiltonian}
H=\sum_{\langle i,j \rangle} J_{i, j} H_{(i,j)}  =\sum_{\langle i,j \rangle}  J_{i, j} \sum_{\mu, \nu}\hat{S}_{i}^{\mu\nu}\hat{S}_{j}^{\nu\mu},
\end{align}
where $J_{i,j}$ are the coupling constants between sites $i$ and $j$. The color indices $\mu,\nu$ take $N$ values (which we may call $A,B,C,\ldots$), and the \SU{N} generators at each site $i=1,\ldots,N_s$ satisfy the commutation relation
\begin{equation} \label{SUN_commut} 
	\big{[}\hat{S}^{\alpha\beta}_{i},\hat{S}^{\mu\nu}_{i} \big{]} = \delta^{\mu\beta}\hat{S}^{\alpha\nu}_{i}-\delta^{\alpha\nu}\hat{S}^{\mu\beta}_{i}.
\end{equation}

When each site hosts one particle in the fundamental irrep of \SU{N}, the local Hilbert space is $N$-dimensional and spanned by $N$ states, one for each color $\mu=A,B,C,\ldots$.
One can then represent the traceless generators with fermionic creation and annihilation operators carrying color and site indices (i.e. for $\mu,\nu=1,\ldots,N$ and $i=1,\ldots,N_s$) as
\begin{equation*}
\hat{S}^{\mu\nu}_{i}= c_i^{\mu\dagger}c_i^{\nu}- \frac{\delta^{\mu,\nu}}{N}.
\end{equation*}
%For $N=2$, calling $A \equiv \, \uparrow$ and $B \equiv \, \downarrow$, one recovers the spin-$1/2$ operators: $\hat{\sigma}^z_j = \frac{1}{2} (\hat{S}^{\uparrow\uparrow}_{j}-\hat{S}^{\downarrow\downarrow}_{j})$, %$\hat{\sigma}^x_j = \frac{1}{2} (\hat{S}^{\uparrow\downarrow}_{j}+\hat{S}^{\downarrow\uparrow}_{j})$
For $N=2$, calling $A \equiv \, \uparrow$ and $B \equiv \, \downarrow$, one recovers the spin-$1/2$ operators: $\hat{S}^z_j = \frac{1}{2} (\hat{S}^{\uparrow\uparrow}_{j}-\hat{S}^{\downarrow\downarrow}_{j})$, $\hat{S}^x_j = \frac{1}{2} (\hat{S}^{\uparrow\downarrow}_{j}+\hat{S}^{\downarrow\uparrow}_{j})$
%and $\hat{\sigma}^y_j = \frac{i}{2} (\hat{S}^{\downarrow\uparrow}_{j}-\hat{S}^{\uparrow\downarrow}_{j})$,
and $\hat{S}^y_j = \frac{i}{2} (\hat{S}^{\downarrow\uparrow}_{j}-\hat{S}^{\uparrow\downarrow}_{j})$,
%so that the two-site interaction term in Eq.~\eqref{hamiltonian} is $H_{(i,j)} =\sum_{\mu, \nu \in \{ \uparrow, \downarrow \} }\hat{S}_{i}^{\mu\nu}\hat{S}_{j}^{\nu\mu} = 2 \hat{\sigma}_i \cdot \hat{\sigma}_j$.
so that the two-site interaction term in Eq.~\eqref{hamiltonian} is $H_{(i,j)} =\sum_{\mu, \nu \in \{ \uparrow, \downarrow \} }\hat{S}_{i}^{\mu\nu}\hat{S}_{j}^{\nu\mu} = 2 \, \hat{\mathbf{S}}_i \cdot \hat{\mathbf{S}}_j$.
For every $N \geq 2$, this interaction takes the form of a permutation operator $P_{i,j}$:
\begin{align}
\label{interaction_ij_oneparticle}
H_{(i,j)}=\sum_{\mu, \nu}\hat{S}_{i}^{\mu\nu}\hat{S}_{j}^{\nu\mu}=P_{i,j}-\frac{1}{N},
\end{align}
where $P_{i,j}$ exchanges the states at sites $i$ and $j$: $P_{i,j} |\gamma\rangle_i \otimes  |\beta\rangle_j=|\beta\rangle_i \otimes  |\gamma\rangle_j$,
for any $\gamma, \beta=A,B,C, ...$.
In the following, the constant $1/N$ is dropped.

\paragraph{Decomposition of the Hilbert space.}

The full Hilbert space $\Box^{\otimes N_s}$ (where $\Box$ stands for the fundamental irrep) can be decomposed as ~\cite{nataf2014,PRBNataf2016,Wan2017}
\begin{equation}
\label{eq: decomposition_hilbert}
\Box^{\otimes N_s} = \underset{\alpha}{\oplus} \overset{d^{\alpha}_N}{\underset{j = 1}{\oplus}} V^\alpha_j,
\end{equation}
where the outer sum runs over all Young diagrams $\alpha$ with $N_s$ boxes and at most $N$ rows.
Each of these Young diagrams (or tableaux) $\alpha=[\alpha_1,\alpha_2,\ldots,\alpha_k]$ (where the row lengths $\alpha_j$ satisfy $\alpha_1\geq\alpha_2\geq\cdots\geq \alpha_k \geq 1$ and $\sum_k \alpha_k=N_s$) labels an irrep of \SU{N} of dimension $d^{\alpha}_N$, an integer that can be calculated from $N$ and the shape $\alpha$ using, e.g., the hook-length formula~\cite{Weyl1925Dec,Robinson1961}.
For instance, $d^{\alpha=[2]}_{N=2} = 3$ corresponds to the three two-site \SU{2} triplets.

\begin{figure}
    \centering\includegraphics[width=.9\linewidth]{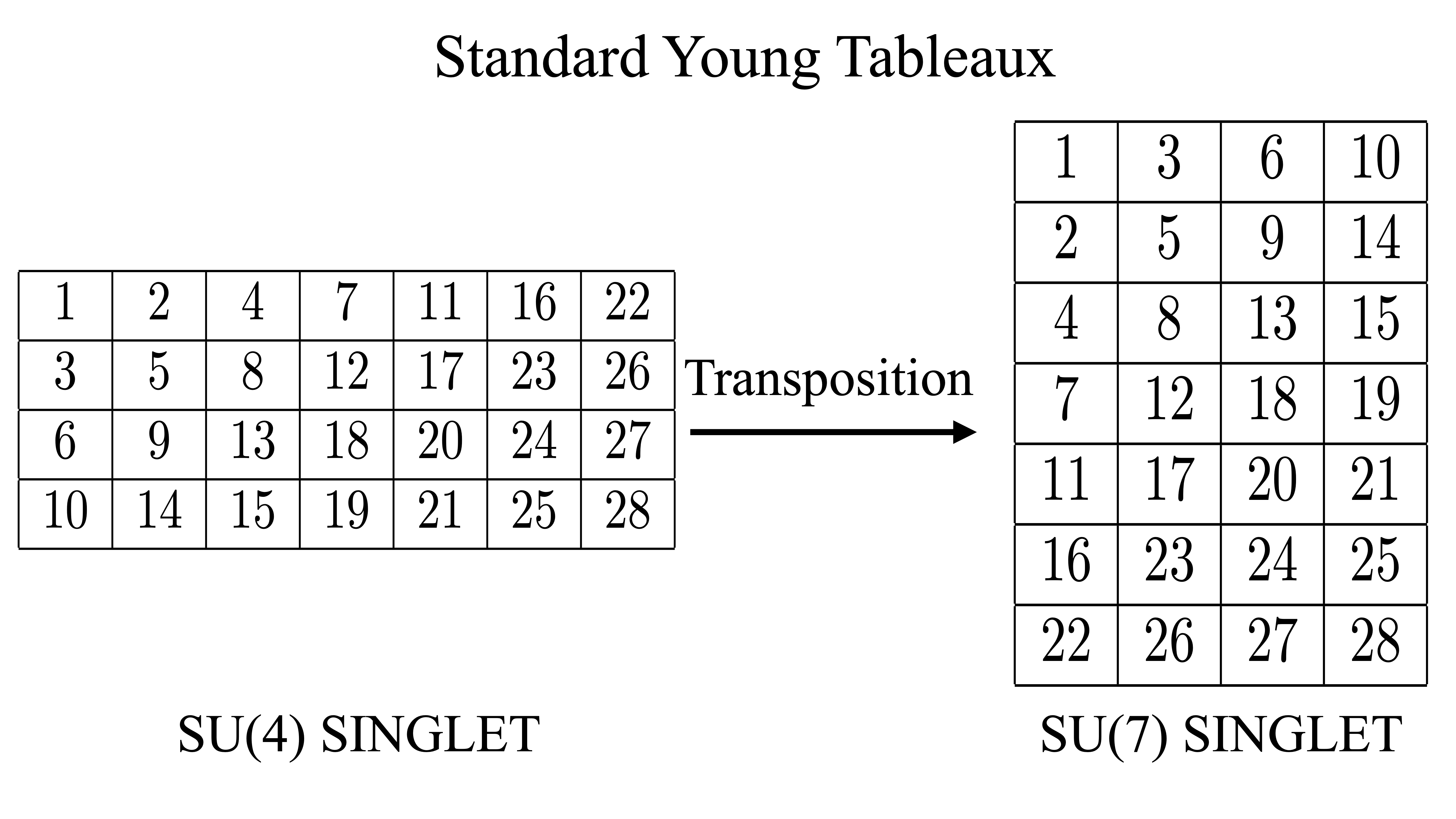}
    \caption{
Left: SYT of shape $\alpha=[7,7,7,7]$, representing a \SU{4} singlet state on $N_s=28$ sites.
Right: SYT for the shape $\alpha=[4,4,4,4,4,4,4]$, representing a \SU{7} singlet state on $N_s=28$ sites. It is obtained from the SYT shown on the left through a \textit{transposition} operation, whereby the fully filled columns (rows) of the \SU{N} tableau are mapped onto the fully filled rows (columns) of the \SU{N_s/N} tableau. 
    }
    \label{SYT_singlets}
\end{figure}

For a given Young diagram $\alpha$, there are $d_N^\alpha$ invariant subspaces $V_j^\alpha$ ($j = 1,\ldots,d_N^\alpha$), 
each invariant under the action of the Hamiltonian $H$ and, more generally, 
under any element of the group algebra of $\mathcal{S}_{N_s}$. 
These subspaces are mutually isomorphic and have the same dimension $f^\alpha$, 
equal to the number of standard Young tableaux (SYTs) of shape $\alpha$.
An SYT is obtained by filling the boxes of $\alpha$ with the integers $1,\ldots,N_s$ such that the entries increase from left to right along each row and from top to bottom along each column.
Figure~\ref{SYT_singlets} shows one SYT of shape $\alpha=[7,7,7,7]$ (left) and one SYT of shape $\alpha=[4,4,4,4,4,4,4]$ (right), 
selected among the $f^{[7,7,7,7]} = f^{[4,4,4,4,4,4,4]} = 13'672'405'890$ possible tableaux.
As a manifestation of Schur--Weyl duality, the same Young diagram $\alpha$ also labels an irrep of the symmetric group $\mathcal{S}_{N_s}$, whose dimension is $f^\alpha$.
A natural basis for this irrep -- and equivalently for each sector $V_j^\alpha$ ($j=1,\ldots,d_N^\alpha$) -- is provided by the complete set of SYTs of shape $\alpha$.
Therefore, the dimension of the full Hilbert space
satisfies $\text{dim} (\Box^{\otimes N_s}) =N^{N_s}=\sum_{\alpha} f^\alpha d^{\alpha}_N$.

The subspaces $V^\alpha_j$ ($j=1,\ldots,d^{\alpha}_N$) are connected by the total lowering and raising operators ($1 \leq \sigma, \mu \leq N$), defined as
\begin{equation}
F^{\sigma, \mu} = \sum_{i=1}^{N_s} c^{\sigma\dagger}_{i} c^{\mu}_i,
\end{equation}
which generalize to \SU{N} the total operators $S^{+} = \sum_{j=1}^{N_s} \hat{S}^+_j = (S^{-})^{\dag}$ and $S^{z} = \sum_{j=1}^{N_s} \hat{S}^z_j $ for \SU{2}.
They also obey the \SU{N} commutation relations in Eq.~\eqref{SUN_commut} and commute with the permutation operators $P_{i,j}$~\cite{Botzung_2023}:
\begin{equation}
\label{commutation_F_E}
[F^{\sigma, \mu}, P_{i,j}] = 0. 
\end{equation}

For the \SU{N} HM Hamiltonian in Eq.~\eqref{hamiltonian}, $d^{\alpha}_N$ therefore appears as a multiplicity (or degeneracy) within the full spectrum.
In particular, when $N$ divides $N_s$ and $\alpha$ is the rectangular shape with $N$ rows and $N_s/N$ columns (see Fig.~\ref{SYT_singlets} for $N=4$ and $N_s=28$) then $d^{\alpha}_N = 1$.
This shape represents the \SU{N} singlet sector, on which the relevant \SU{N} generators act trivially.

The spectrum is obtained by diagonalizing $H$ (seen as an element of the algebra of $\mathcal{S}_{N_s}$) in each $f^\alpha$-dimensional irrep $\alpha$ of $\mathcal{S}_{N_s}$, 
following the rules of the orthogonal representation of $\mathcal{S}_{N_s}$, which we review in the next paragraph.
Before that, let us mention that there exist {\it dual} Hamiltonians, i.e. multilevel (or \SU{N}) Lipkin--Meshkov--Glick models~\cite{Gilmore_1979,
Carrasco_2016,
Carrasco_2017,
Calixto_2021}, written as quadratic polynomials in the $F^{\sigma, \mu}$ and for which the dimension and multiplicity are interchanged, being respectively $d^{\alpha}_N$ and $f^\alpha$.

%Their number is denoted by $f^\alpha$, and they can be  ranked from 1 to $f^{\alpha}$ according to the {\it last letter sequence}: two SYTs $S_r$ and $S_s$ are such that $S_r<S_s$ if the number $N_s$ appears in $S_r$ in a row below the one it appears in $S_s$. If those rows are the same, one looks at the rows of $N_s-1$, etc  (see Fig.\ref{schema_short} b).

\paragraph{Matrix representation of $H$.}

We denote by $\mathcal{H}^{\alpha}$ the matrix representation of $H = \sum_{\langle i,j \rangle} J_{i, j} P_{i,j}$ in the irrep $\alpha$ of $\mathcal{S}_{N_s}$.
It is readily constructed in the SYT basis using Young's orthogonal representation of $\mathcal{S}_{N_s}$. Any transposition $P_{i,j}$ can first be expressed as a product of an \emph{odd} number of adjacent transpositions $P_{k,k+1}$, with $k$ ranging from $\min(i,j)$ to $\max(i,j)-1$.

\begin{figure}
    \centering\includegraphics[width=1\linewidth]{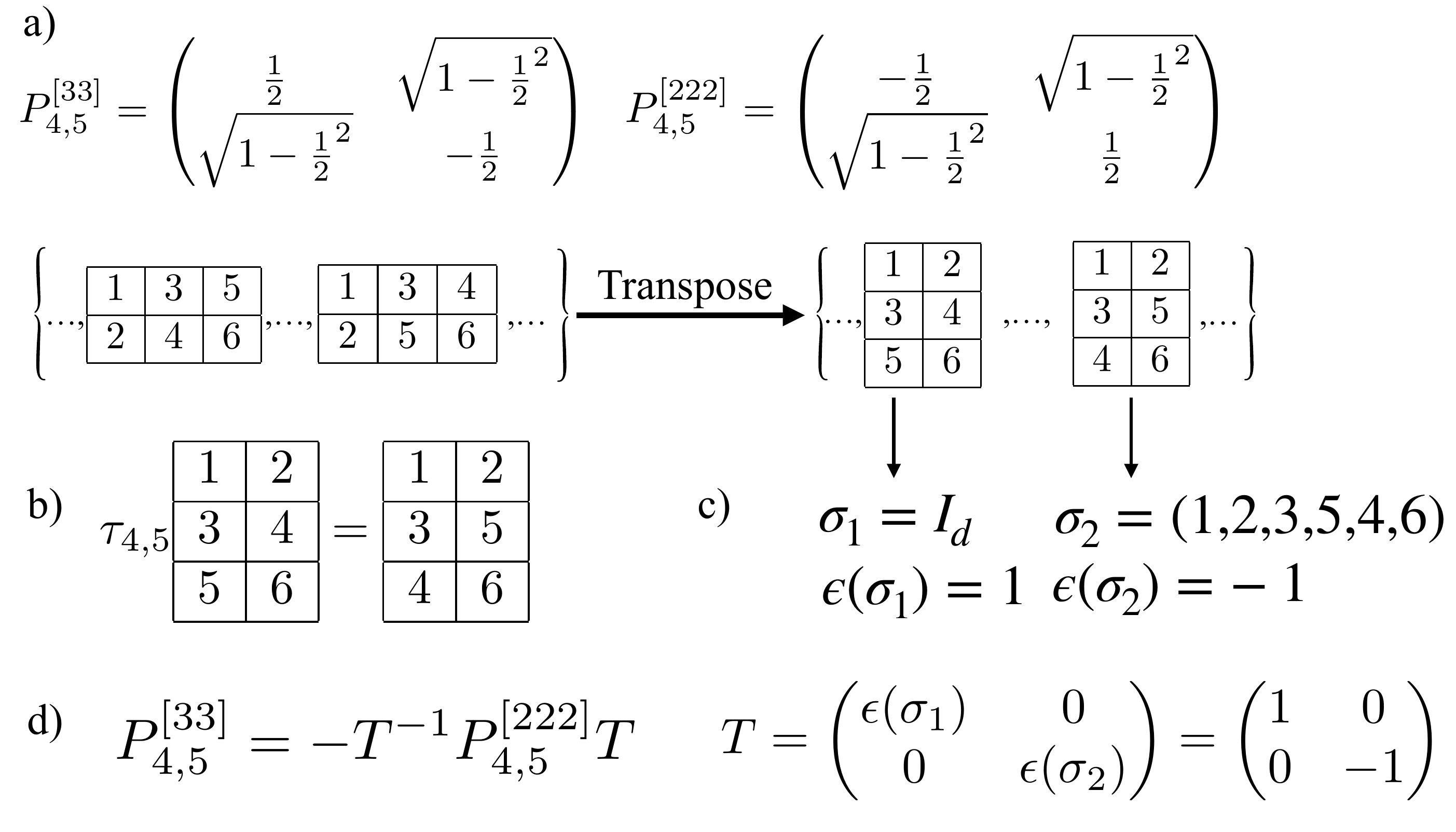}
    \caption{
   (a)~Left: Matrix representation of $P_{4,5}^{[3,3]}$ in the subspace of the two SYTs shown below. The axial distance from $k=4$ to $k+1=5$ in the first SYT is $-2$. Right: After transposition, the resulting SYTs belong to the basis associated with the shape $[2,2,2]$. The corresponding $2\times2$ submatrix of $P_{4,5}^{[2,2,2]}$ coincides with that of $P_{4,5}^{[3,3]}$, except that the signs of the diagonal entries are reversed.
   (b)~Tableau transposition $\tau_{4,5}$ that permutes $4$ and $5$ inside a SYT.
   (c)~To each SYT of a given shape, one can associate a permutation by reading its entries from left to right and top to bottom.
  (d)~The similarity transformation is obtained from the diagonal signature matrix $T$.
    }
    \label{fig:theorem}
\end{figure}

Then, the matrix $P^{\alpha}_{k,k+1}$ describing $P_{k,k+1}$ in the irrep $\alpha$ takes a very simple form in this basis.
Denoting by $\mu_{tu}^{\alpha}(P_{k,k+1})$ the matrix element of $P_{k,k+1}^{\alpha}$ between the SYT $S_t$ and $S_u$, the following cases arise:
if $k+1$ and $k$ are in the same row (resp. column) of the SYT $S_t$, then $\mu^{\alpha}_{tt} (P_{k,k+1})=+1$ (resp. $-1$), and all other matrix elements involving $t$ vanish.
If $k+1$ and $k$ lie in neither the same column nor the same row, and if $S_u$ is the tableau obtained from $S_t$ by interchanging $k$ and $k+1$, then the only non-vanishing matrix elements involving $t$ or $u$ are given by
\begin{align} 
\begin{pmatrix} \label{orthogonal_representation} 
  \mu^{\alpha}_{tt} (P_{k,k+1})&\mu^{\alpha}_{tu} (P_{k,k+1}) \\
   \mu^{\alpha}_{ut} (P_{k,k+1})&\mu^{\alpha}_{uu} (P_{k,k+1}) 
\end{pmatrix}=\begin{pmatrix} -\rho&\sqrt{1-\rho^2} \\ 
  \\ \sqrt{1-\rho^2} &\rho
\end{pmatrix} 
\end{align} 
where $\rho$ is the inverse of {\it the axial distance} from $k$ to $k+1$ in $S_t$, defined by assigning $+1$ (resp. $-1$) to each step made downwards or to the left (resp. upwards or to the right) to reach $k+1$ from $k$.
Figure~\ref{fig:theorem} presents, for two pairs of SYTs of shapes $\alpha=[3, 3]$ and $\alpha=[2, 2, 2]$, the corresponding $2\times2$ submatrices arising from Eq. \eqref{orthogonal_representation}.

$\mathcal{H}^{\alpha}$ is then obtained as a linear superposition of products of matrices of the preceding form.

\subsection{Mapping Theorem between the singlet spectra of the \SU{N} and the \SU{N_s/N} Heisenberg models}

We are now in a position to state the mapping theorem. Let
\begin{equation}
H = \sum_{\langle i,j \rangle} J_{i, j} P_{i, j}
\end{equation}
be the HM Hamiltonian on an $N_s$-site lattice, 
with $p=N_s/N \in \mathbb{N}^*$ and the fundamental irrep of \SU{N} at each site,
and $P_{i, j}$ the permutation operator between sites $i$ and $j$ (see Eq. \ref{interaction_ij_oneparticle} for the connection with the local \SU{N} generators).

 Let $\epsilon_1 \leqslant \epsilon_2 \leqslant \cdots \leqslant \epsilon_{f^{[p,\ldots,p]}}$ denote the full spectrum of $H$ in the global \SU{N} singlet sector.
 Then, the spectrum of $H$ on the same $N_s$-site lattice for \SU{N_s/N} spins in the global \SU{N_s/N} singlet sector is $-\epsilon_{f^{[p,\ldots,p]}} \leqslant \cdots \leqslant -\epsilon_2 \leqslant -\epsilon_1$.

{\it Proof:}

In fact, one can prove a stronger result. Let us introduce the {\it transposition} map, which transforms columns (possibly containing integers arranged in ascending order from top to bottom) into rows (containing the same integers arranged in ascending order from left to right).

Denoting by $\bar{\alpha}$ the transpose of the Young diagram $\alpha=[\alpha_1,\alpha_2,\cdots,\alpha_k]$ (i.e. $\bar{\alpha}$ is the Young diagram whose columns have lengths $\alpha_1,\alpha_2,\cdots,\alpha_k$), there exists a signature matrix $T$ (namely, a diagonal matrix whose entries are $\pm 1$) such that
\[
\mathcal{H}^{\bar{\alpha}}=-T^{-1}\mathcal{H}^{\alpha}T.
\]

We first note that the transposition establishes a one-to-one correspondence between the full basis of SYTs of shape $\alpha$ and that of shape $\bar{\alpha}$.
In these respective bases, the matrix elements of $P_{k,k+1}$ in the irrep $\bar{\alpha}$ (denoted $P^{\bar{\alpha}}_{k,k+1}$) coincide with those of $P^{\alpha}_{k,k+1}$, 
except for the diagonal entries, whose signs are reversed (see Fig.~\ref{fig:theorem}a), since rows and columns are exchanged under transposition.
Extracting an overall minus sign, we seek a similarity transformation that flips the sign of every off-diagonal entry of all matrices $P^{\bar{\alpha}}_{k,k+1}$ ($k=1,\ldots,N_s-1$).

Consider the graph whose vertices are all SYTs of shape $\bar{\alpha}$ and whose edges join pairs related by a {\it tableau transposition} $\tau_{k,k+1}$. 
Such a transposition interchanges the entries $k$ and $k+1$ inside a SYT and thereby generates another SYT (see Fig.~\ref{fig:theorem}(b)), 
so that $P_{k,k+1}$ is a linear combination of the identity and $\tau_{k,k+1}$ (cf Eq. \ref{orthogonal_representation}).
To each SYT, one can associate a permutation by reading its entries from left to right and top to bottom \cite{LLM_contribution}, as illustrated by Fig.~\ref{fig:theorem}(c) .
The signature of this permutation changes sign under each tableau transposition, making the graph bipartite.

The diagonal entries of $T$ are precisely given by the signatures of the corresponding SYTs. As a consequence,
\[
P^{\alpha}_{k,k+1}
=
-T^{-1}P^{\bar{\alpha}}_{k,k+1}T,
\qquad
\forall\, k=1,\ldots,N_s-1,
\]
and therefore, since $H$ is a linear superposition of products containing an odd number of successive transpositions,
\[
\mathcal{H}^{\alpha}
=
-T^{-1}\mathcal{H}^{\bar{\alpha}}T.
\]
In particular, when $\alpha$ corresponds to the \SU{N} singlet sector on $N_s$ sites (i.e. when $\alpha$ is a rectangular Young diagram with $N$ rows and $p=N_s/N$ columns), 
its transpose $\bar{\alpha}$ corresponds to the $\mathrm{SU}(p) = \mathrm{SU}(N_s/N)$ singlet sector, which completes the proof.

\section{Indexing, storing, and using SYTs}
\label{tech_aspec}

%\subsection{Indexing SYTs\label{sec::indexing_SYTs}}

To perform ED while exploiting the full \SU{N} symmetry, it is necessary to index an SYT within the basis of all $f^{\alpha}$ SYTs associated with an irrep $\alpha$ of \SU{N}.
The basis of SYTs can be ordered in the so-called increasing or decreasing order of the last letter order sequence \cite{nexytb}. 
In this ordering, two SYTs are compared as $Y^{\alpha}_{m} > Y^{\alpha}_{m'}$ if the number $N_s$ appears in $Y^{\alpha}_{m}$ in a row below the one it appears in $Y^{\alpha}_{m'}$. 
If $N_s$ appears in the same row in $Y^{\alpha}_{m}$ and $Y^{\alpha}_{m'}$, one looks at $N_s-1$, etc. 
This order provides a natural ordering of the basis, and will prove to be extremely useful.

Indexing an SYT, namely finding its position in the ordered basis of all SYTs, can be done in several ways, which were already explored and used in Ref~\cite{nataf2014}. 
At one end of the spectrum, one can adopt a basis-free approach that requires only $\mathcal{O}(1)$ additional memory. 
A SYT is then identified with a path in the branching graph obtained by recursively removing corner boxes, to which standard ranking and unranking procedures can be applied~\cite{Wilf1977,RaoShankar2015}.
The memory benefit of this approach is to be balanced with its computational cost. At the other end of the spectrum, 
one can store the basis of SYTs entirely, and use a binary search to index SYTs. 
This is by far the method of choice, but storing billions of SYTs may become a challenge. 
A hybrid approach consists in constructing a set of partial lookups that implement the branching 
of the canonical group chain of $\mathcal{S}_{N_s}$ (i.e. $\mathcal{S}_1 \subset \mathcal{S}_2 \subset \cdots \subset \mathcal{S}_{N_s}$), namely storing all sub-SYTs with $n_1<N_s$ boxes. 
This allows a fast binary search for the first $n_1$ boxes while keeping the ``high'' part of the basis elements to be indexed as in the basis-free approach.
This hybrid approach has the benefit of being versatile since the memory usage can be tuned by changing 
the number $n_1$ of looked-up boxes.

%\subsection{Storing SYTs}

\begin{table*}[t]
\centering
\renewcommand{\arraystretch}{2.0}  % <-- increases all row heights
\begin{tabular}{c|cccccc}
$m$ & 1 & 2 & 3 & 4 & 5 & \makecell{Increasing order} \\
\hline
$Y^{\alpha}_m$ & 
$\shsc{4pt}{1}{\ytableaushort{14,25,36}}$ & 
$\shsc{4pt}{1}{\ytableaushort{13,25,46}}$ & 
$\shsc{4pt}{1}{\ytableaushort{12,35,46}}$ & 
$\shsc{4pt}{1}{\ytableaushort{13,24,56}}$ & 
$\shsc{4pt}{1}{\ytableaushort{12,34,56}}$ & 
\multirow{2}{*}[-2ex]{\makecell{Decreasing \\order\\ of LLOS}}\\[2ex]
\cline{1-6}
\makecell{Yamanouchi \\symbol} & 
$[1, 2, 3, 1, 2, 3]$ & 
$[1, 2, 1, 3, 2, 3]$ & 
$[1, 1, 2, 3, 2, 3]$ & 
$[1, 2, 1, 2, 3, 3]$ & 
$[1, 1, 2, 2, 3, 3]$ & \\
\hline
bit-word & 
$100100100100$ & 
$100110000100$ & 
$100110010000$ & 
$101001000100$ & 
$101001010000$ & 
\makecell{Increasing\\page order}\\
\hline
value & 
2340 & 
2436 & 
2448 & 
2628 & 
2640 & 
\makecell{Increasing\\value order}\\
\end{tabular}
\caption{The five SYTs for the $N_s=6$ singlet irrep of \SU{3}, ordered in descending last-letter order. The bit word of each SYT is obtained by subtracting 1 from each entry of the Yamanouchi symbol, converting each result to its $\eta=2$-bit pattern, and reversing the six two-bit entries. The row position of the first box, which is always in the first row, then appears in the two least significant bits, and that of the last box in the two most significant bits.}
\label{tab:syts}
\end{table*}

\begin{figure}
    \centering
    \includegraphics[width=1.00\linewidth]{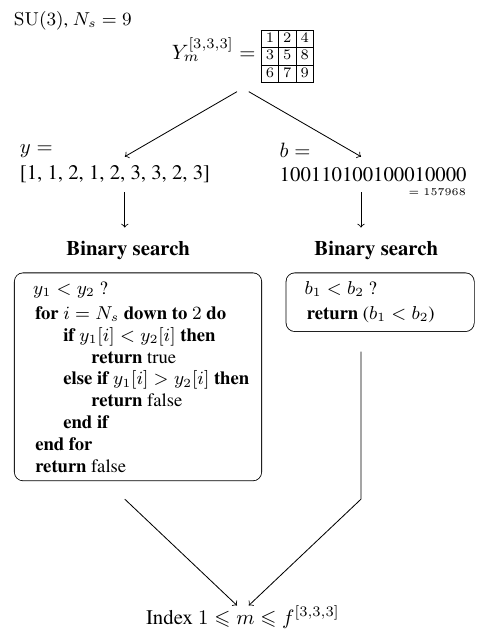}
    \caption{Indexing procedure for SYTs. When using Yamanouchi symbols to represent SYTs (left path), the weak ordering operator compares box positions until it finds a difference. This requires up to $N_s-1$ comparisons, since the $1$ is always placed in the top-left box. By contrast, the bit-level storage strategy naturally implements the inverse of the last-letter order sequence, and the weak ordering operator requires a single comparison (right path).}
    \label{fig:indexing_SYT}
\end{figure}

A SYT $Y^{[\nu]}_m$ associated with a Young diagram $\nu$ containing $N_s$ boxes can be stored as an ordered sequence of the $N_s$ row positions of its entries, which corresponds to the path in the canonical group chain followed to reach $\nu$. This is the Yamanouchi representation \cite{Yamanouchi1937}.
In practice, one may store a Yamanouchi symbol as an array $[y_1,y_2,\ldots,y_{N_s}]$, with $1 \leqslant y_i \leqslant n_r$ and $n_r$ the number of rows in the Young diagram, 
as illustrated in Tab.~\ref{tab:syts} and Fig.~\ref{fig:indexing_SYT}. 
Allocating one byte per entry requires $N_s$ bytes per SYT, and thus $N_s f^{\nu}$ bytes for all $f^{\nu}$ SYTs. 
For instance, the 13.67 billion SYTs of the singlet irrep of \SU{4} with $N_s=28$ would require over 382~GB of memory, 
namely 3.5 times more than a real double-precision Lanczos vector.
However, since the experimentally relevant values of $N$ are moderate, $2 \leqslant N \leqslant 10$, 
and the relevant irreps of $\mathcal{S}_{N_s}$ have $n_r \leqslant N$, at most 4 of the 8 bits per box are needed. 
This storage strategy therefore uses at least twice as many bits as strictly necessary. We now describe a memory-optimal and computationally efficient alternative.

In order to reduce the total memory usage, but still benefit from the possibility of using a full binary search to index SYTs, we pack each SYT into a single word. 
Indeed, for an irrep of \SU{N} with $n_r \leqslant N$ rows (when keeping columns with $N$ boxes), one needs $\eta = \lceil \log_2 n_r \rceil$ bits per box (indexing rows from $0$ to $n_r - 1$).
Thus, an SYT can be stored using $\eta N_s$ bits\footnote{Since the tableaux are filled in the \emph{standard} way, one would strictly need only $\eta (N_s-1)$ bits to store an SYT associated with an $N_s$-box irrep. In practice, we still store the $\eta$ zero bits associated with the first box to avoid an extra addition when retrieving or setting box positions. This could be adapted in a limiting case, for instance for a \SU{3} model with $N_s=33$.}.
For the $N_s=28$ \SU{4} system, we can thus pack each SYT into a single 8-byte integer. 
This reduces the memory cost of storing the entire SYT basis from 382~GB to 109~GB, the equivalent of a Lanczos vector.

In addition to the memory gain, this packed approach naturally encodes the last-letter order sequence (in reverse order). Comparing two SYTs then reduces to a single comparison of two 8-byte words, instead of up to $N_s-1$ comparisons of 1-byte words in the naive storage scheme. This is illustrated in Table~\ref{tab:syts} and Fig.~\ref{fig:indexing_SYT}, respectively.

Beyond the spectral mapping established in Sec.~\ref{theorem}, we observe that, with the approach described above, storing the 13.67 billion SYTs of \SU{7} with $N_s=28$ would require more memory than for \SU{4}: $\eta = \lceil \log_2 7\rceil = 3$ bits are needed per box, requiring one 16-byte word for each SYT.
This extra cost is easily avoided by storing column positions instead of row positions: $\eta = \min\{ \lceil \log_2 n_r\rceil, \lceil \log_2 n_c\rceil \}$, where $n_c$ is the number of columns in the Young diagram, $n_r \leqslant N$, and $n_r = N$ for most irreps of interest, since columns of $N$ boxes are preserved.

This efficient SYT representation is fully compatible with the basis-free and partial-lookup approaches discussed above. 
It significantly reduces memory usage and allows us to store the full basis even for the largest systems. 
We implemented it as a C++ template class using bit-level operations, which can directly replace the naive storage scheme~\cite{gozel_suned_2026}. 
For moderately sized irreps, indexing the entire SYT basis is about $6\times$ faster with the bit-level representation.

The packed representation supplies the indexing and comparison operations required by the Lanczos solver while keeping the full basis in memory. The distributed-memory implementation, the optimization of lattice-site labels, and performance benchmarks are described in Appendix~\ref{app:computational_details}. 

We have decided to apply these technical improvements of the SYT based ED algorithm to the study of the \SU{4} and \SU{7} HM on the TL.

\section{\SU{4} and \SU{7} HM on the triangular lattice}
\label{LRO}
The ground state of the nearest-neighbor \SU{4} Heisenberg antiferromagnet on the
TL, with a fundamental irrep on each site, remains debated.
Early flavor-wave and ED work indicated the instability of
four-sublattice color order (4SL), while a variational approach proposed a resonating
plaquette liquid without four-site plaquette long-range order~\cite{mambrini2003}.
Subsequent DMRG and field-theory work suggested a gapless spin-orbital liquid
with a parton Fermi surface coupled to a U(1) gauge field~\cite{Keselman_2020}.
Other DMRG calculations found bond order at the Heisenberg point~\cite{zhang2021},
while large-$N$ mean-field theory favored four-site valence clusters~\cite{Yao2021}.
More recent DMRG and variational Monte Carlo studies supported a common
critical stripy scenario~\cite{Jin_2022,Zhang2024}.
This state preserves SU(4) symmetry but doubles the unit cell and selects a
preferred lattice direction, hence its designation as a nematic spin-orbital liquid.
Entanglement scaling along the stripes suggests $\mathrm{SU}(4)_1$ criticality with $c\simeq3$~\cite{Jin_2022}.
By contrast, recent parton mean-field calculations~\cite{Kim2026} and
matrix-product-state studies of the undoped $t$--$J$ model on cylinders~\cite{Kadow2026}
favor a plaquette crystal with long-range spatial order.
This ordered scenario differs from the resonating plaquette liquid proposed in Ref~\cite{mambrini2003}.
The competition between critical stripes and plaquette order thus remains
unresolved in the two-dimensional limit.

\begin{table}[t]
\caption{\label{tab:triangular_clusters}
Periodic triangular clusters used in the ED calculations, ordered by increasing
number of sites. In the $\mathbf t_{1,2}$ columns, $(m,n)$ denotes the cluster
translation $m\mathbf a_1+n\mathbf a_2$, with $\mathbf a_1=(1,0)$ and
$\mathbf a_2=(1/2,\sqrt{3}/2)$ as in Fig.~\ref{figJ2c}(a). A superscript
$\Diamond$ ($\star$) on $N_s$ denotes compatibility with four-sublattice
(seven-sublattice) order. The four-sublattice ordering lattice is generated by
$2\mathbf a_1$ and $2\mathbf a_2$. For the seven-sublattice domain of
Fig.~\ref{fig:corr_su7}(c), one may take
$\mathbf T_1=3\mathbf a_1-\mathbf a_2$ and
$\mathbf T_2=-2\mathbf a_1+3\mathbf a_2$; the reflected domain is generated by
$\mathbf T'_1=-\mathbf a_1+3\mathbf a_2$ and
$\mathbf T'_2=3\mathbf a_1-2\mathbf a_2$ (Fig.~\ref{fig:corr_su7}(b)).}
\begin{ruledtabular}
\renewcommand{\arraystretch}{1.15}
\setlength{\tabcolsep}{4pt}
\begin{tabular}{cccc}
$N_s$ & $\mathbf t_1$ & $\mathbf t_2$ & point group \\
\hline
$12^{\Diamond}$ & $(2,2)$ & $(2,-4)$ & $D_6$ \\
$14^{\star}$ & $(1,2)$ & $(5,-4)$ & $C_2$ \\
$14$ & $(1,3)$ & $(4,-2)$ & $C_2$ \\
$16^{\Diamond}$ & $(4,0)$ & $(0,4)$ & $D_6$ \\
$20$ & $(1,5)$ & $(4,0)$ & $C_2$ \\
$20$ & $(2,-4)$ & $(3,4)$ & $D_2$ \\
$20^{\Diamond}$ & $(2,2)$ & $(4,-6)$ & $D_2$ \\
$21^{\star}$ & $(1,4)$ & $(5,-1)$ & $C_6$ \\
$21$ & $(2,-5)$ & $(3,3)$ & $D_2$ \\
$24$ & $(1,4)$ & $(5,-4)$ & $D_2$ \\
$24^{\Diamond}$ & $(2,4)$ & $(4,-4)$ & $C_2$ \\
$28^{\Diamond\star}$ & $(2,4)$ & $(6,-2)$ & $C_6$ \\
\end{tabular}
\end{ruledtabular}
\end{table}

\subsection{Long-range order in the \SU{4} HM with next-nearest-neighbor interactions}
\label{LRO4}

\begin{figure}[t]
    \centering
    \includegraphics[width=\linewidth]{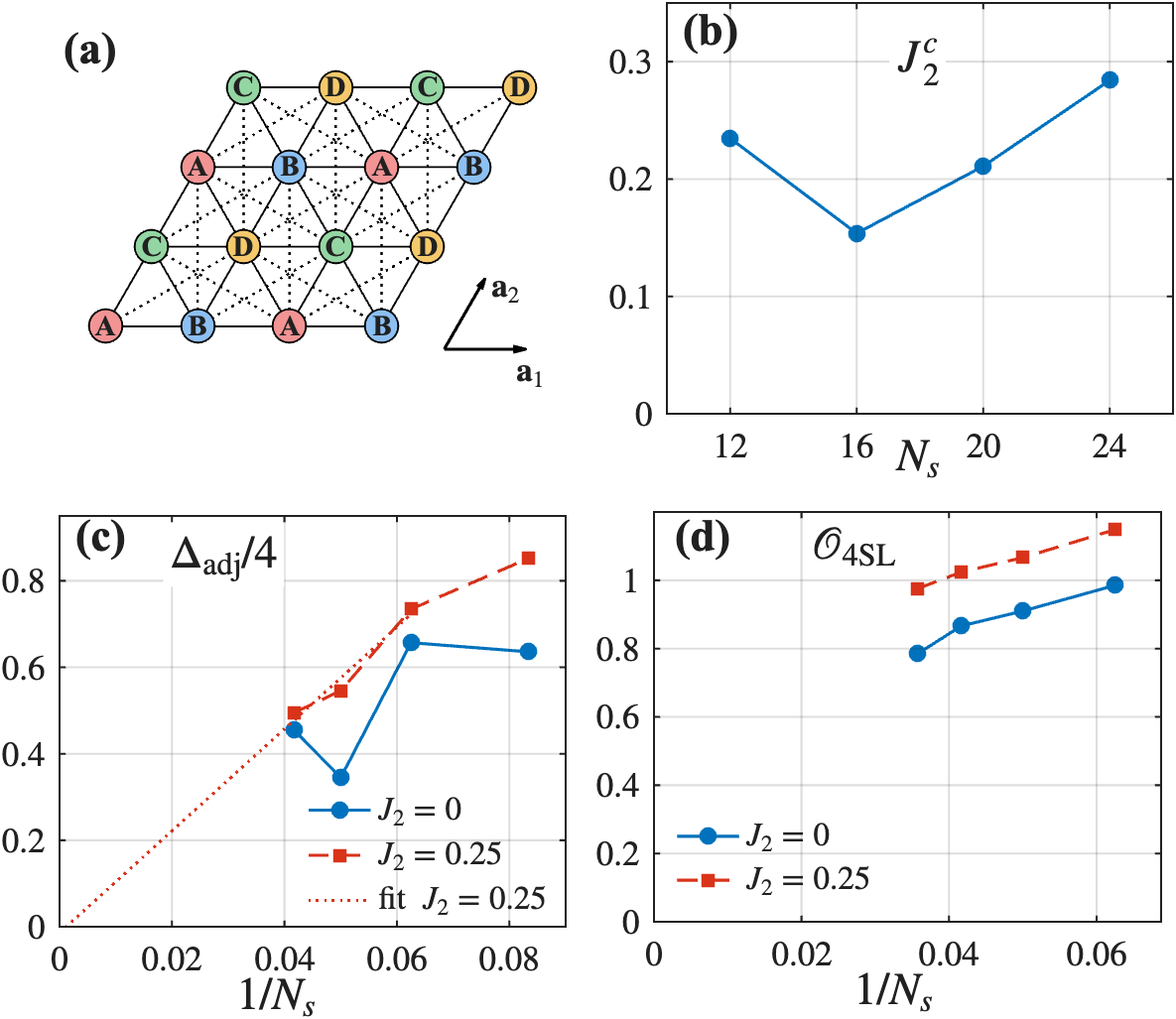}
    \caption{Finite-size diagnostics of four-sublattice order (4SL) in the
    triangular-lattice \SU{4} HM, with $J_1=1$.
    (a)~4SL coloring. Solid and dotted segments denote NN and NNN
    bonds, respectively.
    (b)~Finite-size crossing scale $J_2^c$, defined by the crossing of the
    lowest adjoint gap with the relevant singlet gap and obtained by linear
    interpolation between consecutive sampled values of $J_2$. For $N_s=20$,
    the second singlet excitation is used because the first becomes
    quasi-degenerate with the ground state near $J_2=0$.
    (c)~Tower-of-states slope estimator
    $\Delta_{\mathrm{adj}}/C_2(\mathrm{adj})=\Delta_{\mathrm{adj}}/4$.
    The dotted line is a linear fit to the $J_2=0.25$ data for
    $N_s=16,20,24$, extended over $0\leq 1/N_s\leq1/16$.
    (d)~4SL correlation contrast $\mathcal O_{4\mathrm{SL}}$ (cf Eq.~\eqref{eq:o4sl}).
    Periodically identified directions are counted with their multiplicity.}
    \label{figJ2c}
\end{figure}

We supplement the nearest-neighbor exchange, with coupling strength $J_1$, by a next-nearest-neighbor (NNN)
term with coupling strength $J_2$  (cf Fig.~\ref{figJ2c} a),
\begin{equation}
H=J_1\sum_{\langle i,j\rangle}P_{i,j}
  +J_2\sum_{\langle\!\langle i,j\rangle\!\rangle}P_{i,j},
\label{eq:su4_j1j2}
\end{equation}
and set $J_1=1$ hereafter. The linear flavor-wave theory (LFWT) \cite{Papanicolaou_1984,Papanicolaou1988} of
Ref~\cite{mambrini2003} found that the 4SL state is destabilized at $J_2=0$
by lines of zero modes and divergent quantum fluctuations, whereas a finite
$J_2$ removes this accidental degeneracy and stabilizes 4SL order. That work
estimated a transition near $J_2/J_1\simeq0.12$ and observed the associated
low-energy structure by ED on clusters of at most 16 sites. The larger Hilbert
spaces accessible here allow us to study this mechanism up to $N_s=28$.
The cluster geometries are listed in Table~\ref{tab:triangular_clusters}. A
cluster is 4SL-compatible when both of its torus translations are integer
combinations of $2\mathbf a_1$ and $2\mathbf a_2$ (cf Fig.~\ref{figJ2c} a); such geometries are marked
by a diamond throughout the table.

The NNN perturbation is also physically motivated by the \SU{4}
Fermi--Hubbard model with one particle per site on average (one-quarter filling). ED at
intermediate interaction strength revealed robust 4SL correlations in that
model~\cite{Botzung_2023_PRL}. Beyond the leading Heisenberg exchange, a
strong-coupling expansion generates fourth-order terms of order $t^4/U^3$,
including further-neighbor exchange together with ring and other multispin
processes~\cite{Takahashi_1977,Boos_2020}. Equation~(\ref{eq:su4_j1j2}) should
therefore be viewed as the minimal deformation that isolates the effect of the
NNN exchange, rather than as the complete fourth-order Hubbard Hamiltonian.

\begin{figure}[t]
    \centering
    \includegraphics[width=\linewidth]{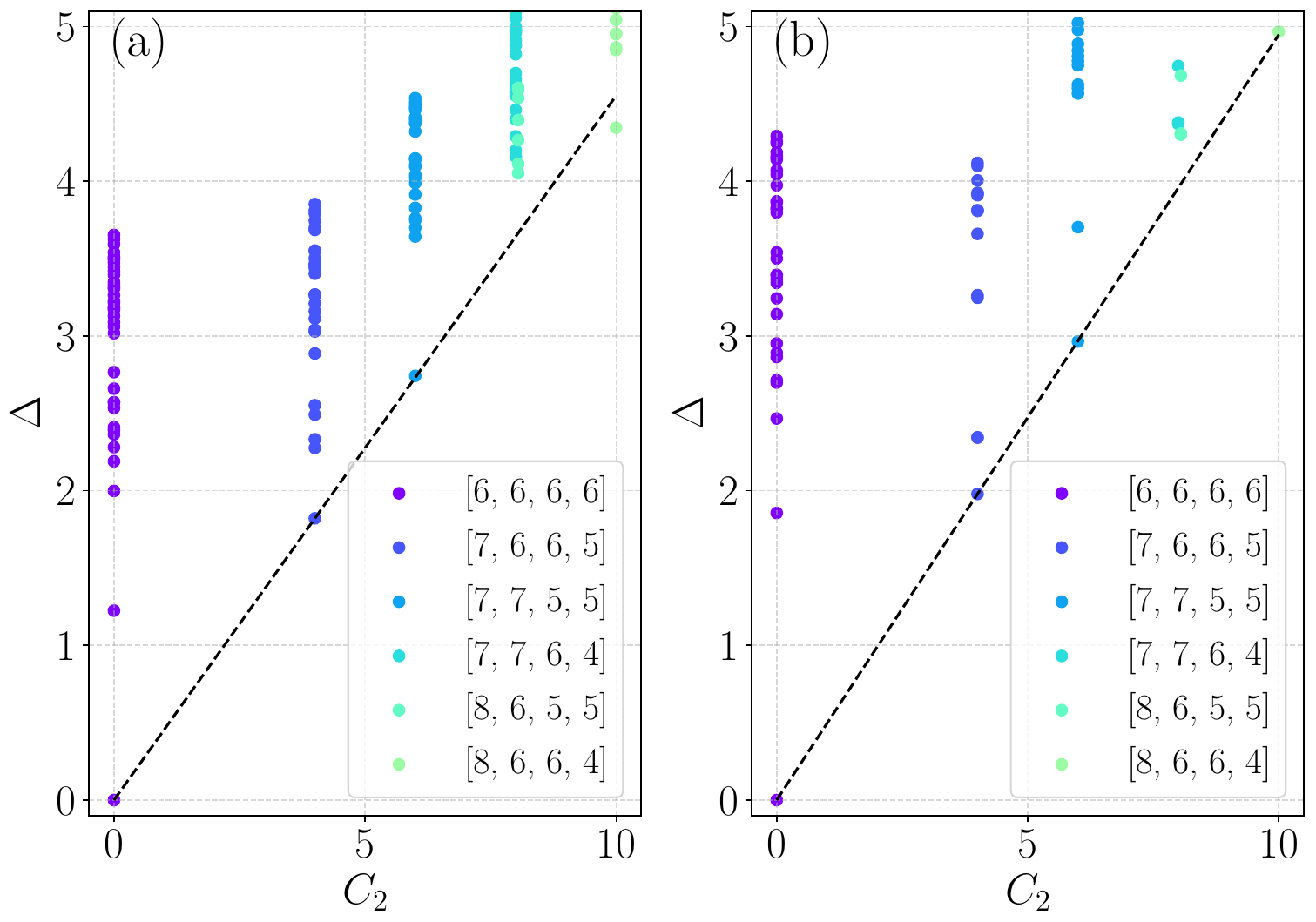}
    \caption{Anderson tower of states on the 4SL-compatible $N_s=24$ cluster.
    Excitation energies are plotted against the quadratic \SU{4} Casimir for
    (a)~$J_2=0$ and (b)~$J_2=0.25$. Colors distinguish the irreps
    $[6,6,6,6]$, $[7,6,6,5]$, $[7,7,5,5]$, $[7,7,6,4]$,
    $[8,6,5,5]$, and $[8,6,6,4]$. The black dashed line indicates the
    form $\Delta_\lambda\propto C_2(\lambda)$ (cf Eq.~\eqref{eq:rotor}).}
    \label{fig:res:2}
\end{figure}

Spontaneous breaking of a continuous symmetry on a finite cluster is encoded
in an Anderson tower of states~\cite{Lecheminant_1994}. Its lowest levels obey
the quantum-rotor form
\begin{equation}
\Delta_\lambda\equiv E_\lambda-E_0
   \simeq A(N_s)C_2(\lambda),\qquad A(N_s)\propto N_s^{-1},
\label{eq:rotor}
\end{equation}
where $C_2(\lambda)$ is the quadratic \SU{4} Casimir \cite{schellekens}. Figure~\ref{fig:res:2}
compares the low-energy spectra of the same 4SL-compatible 24-site cluster at
$J_2=0$ and $J_2=0.25$. The latter exhibits a markedly clearer low-lying
branch, approximately linear in $C_2$.

\begin{figure}[t]
    \centering
    \includegraphics[width=\linewidth]{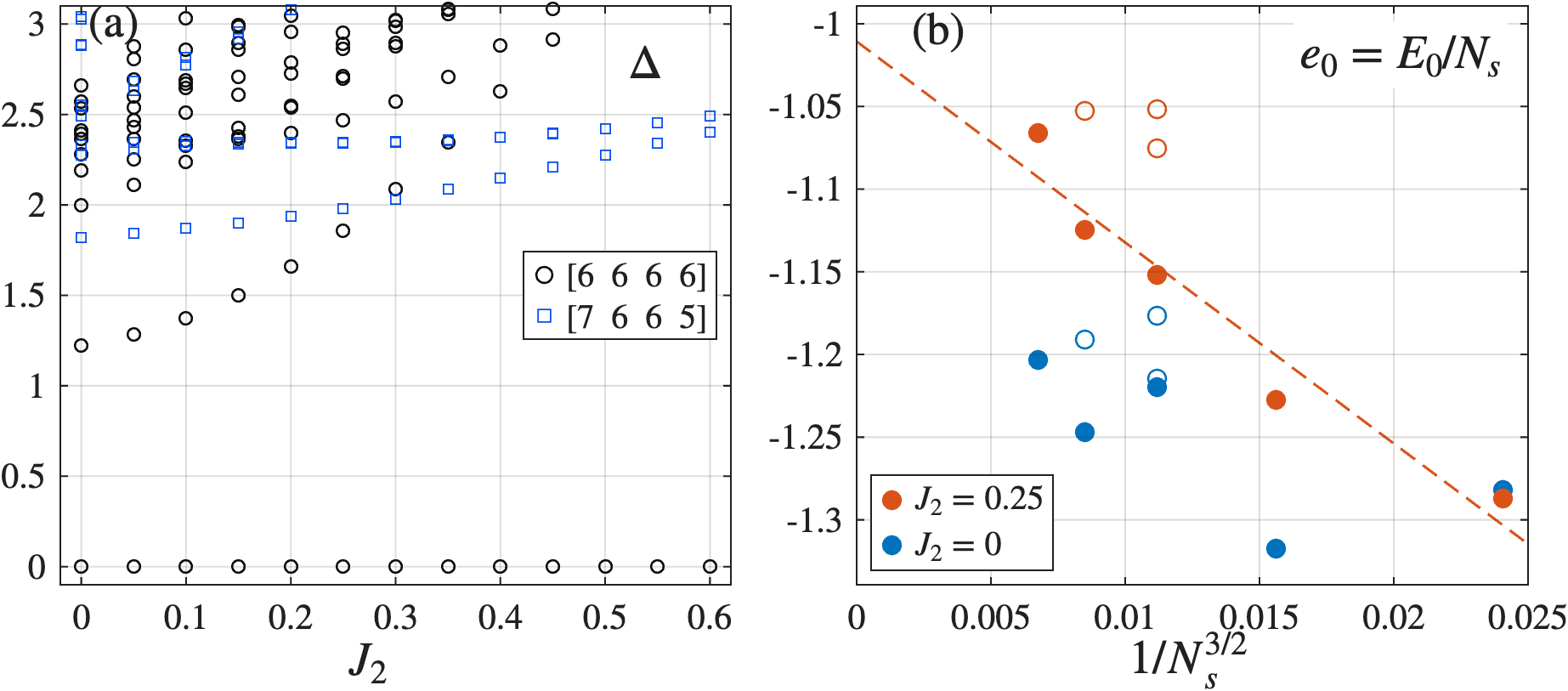}
    \caption{Low-energy spectrum and ground-state energy of the triangular
    \SU{4} model.
    (a)~Excitation energies $\Delta=E-E_0$ on the 4SL-compatible $N_s=24$
    cluster as functions of $J_2$. Open black circles and blue squares denote
    levels in the singlet $[6,6,6,6]$ and adjoint $[7,6,6,5]$ irreps,
    respectively.
    (b)~Ground-state energy per site versus $1/N_s^{3/2}$ for $J_2=0$ and
    $J_2=0.25$. Filled and open symbols denote clusters compatible and
    incompatible with 4SL order, respectively (cf Tab.~\ref{tab:triangular_clusters}). The dashed line is a linear
    fit to the compatible $J_2=0.25$ clusters.}
    \label{fig:res:1}
\end{figure}

This spectral reorganization motivates a finite-size boundary for the onset
of the 4SL regime. We define its position, $J_2^c(N_s)$, by the crossing of the
lowest adjoint gap with the relevant singlet gap. It is a useful
pseudocritical marker, rather than a thermodynamic phase boundary. The
individual levels underlying the $N_s=24$ crossing are displayed in
Fig.~\ref{fig:res:1}(a). The singlet levels belong to $[6,6,6,6]$ and the
adjoint levels to $[7,6,6,5]$. Figure~\ref{figJ2c}(b) then collects the
resulting $J_2^c(N_s)$ for the different cluster sizes. The values remain of
order $0.15$--$0.28$ for $N_s=12$--24 and are nonmonotonic; they therefore
show no systematic drift to zero, but do not support a controlled
extrapolation of $J_2^c$ either.

The collapse of the tower levels can be followed through their slope. Since
$C_2(\mathrm{adj})=4$, the ratio $\Delta_{\mathrm{adj}}/4$ provides the slope
estimator shown in
Fig.~\ref{figJ2c}(c). At $J_2=0.25$, a linear fit to $N_s=16,20,24$ extrapolates
close to zero at $1/N_s=0$, as required for the collapse of the tower levels
in the thermodynamic limit. The extrapolation is based on only three sizes,
but it clearly distinguishes the ordered regime from the less regular
nearest-neighbor data.

A second finite-size signature follows from the ground-state energy. In two
dimensions, linearly dispersing Goldstone modes yield the leading correction
$e_0(N_s)=e_0(\infty)+b/N_s^{3/2}+\cdots$.  Indeed, for a Goldstone
branch $\omega_a(\mathbf q)\simeq c_a|\mathbf q|$ and a smallest nonzero
momentum $q_{\min}\sim N_s^{-1/2}$, its infrared contribution has
the dimensional estimate
\begin{equation}
\delta e_0\sim\sum_a\int_{|\mathbf q|\lesssim q_{\min}}
\frac{\mathrm{d}^2q}{(2\pi)^2}\,\omega_a(\mathbf q)
\propto q_{\min}^3\propto N_s^{-3/2}.
\label{eq:goldstone_energy_scaling}
\end{equation}

As shown in
Fig.~\ref{fig:res:1}(b), the 4SL-compatible clusters at $J_2=0.25$ follow this
form much more coherently than the incompatible geometries. This scaling is
consistent with a two-dimensional N\'eel state, although it is not by itself
a proof of LRO~\cite{Lecheminant_1994,Sorella_1999}.

\begin{figure*}[t]
    \centering
    \includegraphics[width=\textwidth]{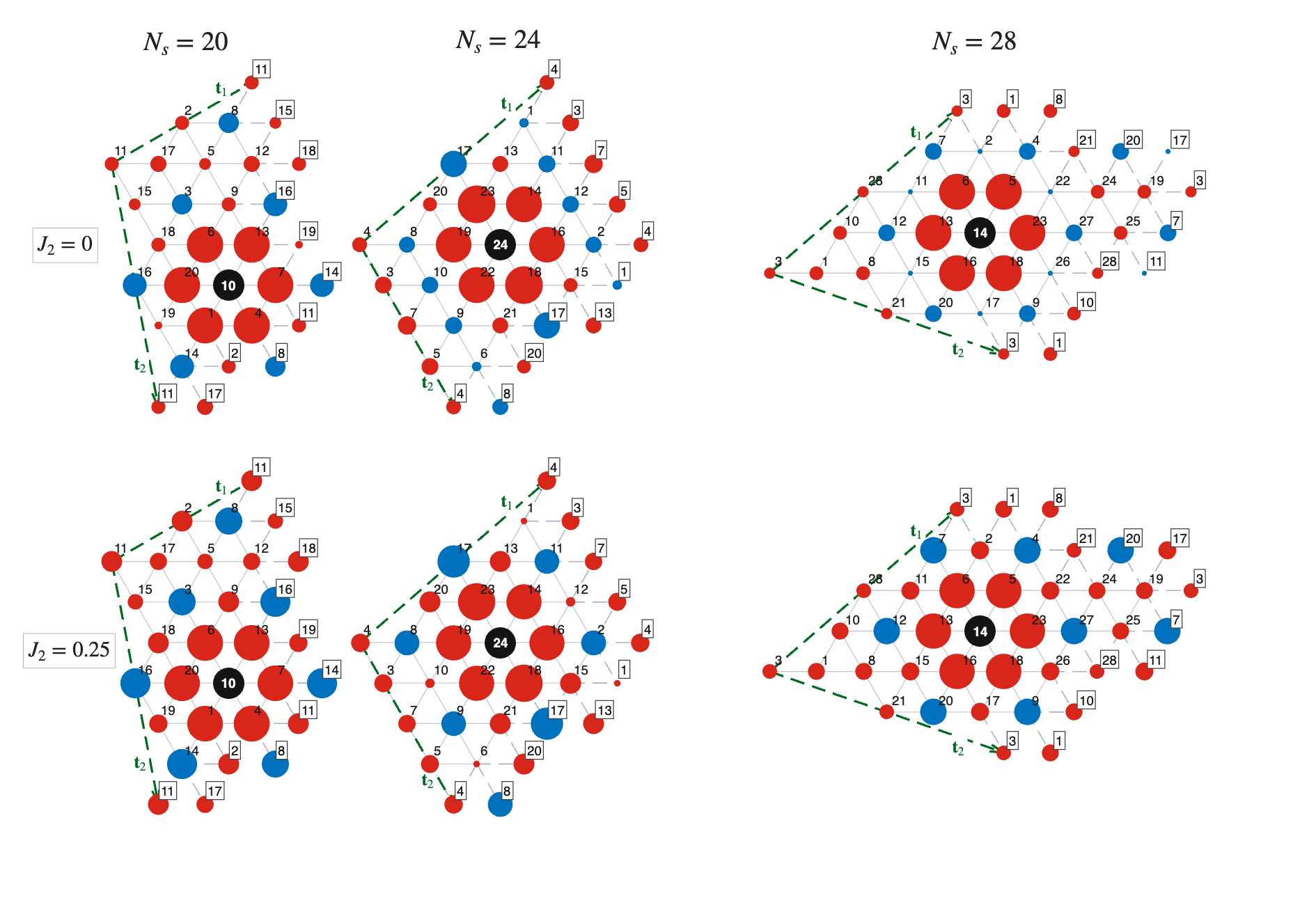}
    \caption{Real-space permutation correlations $\langle P_{j_{\rm ref},j}\rangle-\frac14$ in the \SU{4} ground state.
    Columns correspond to the 4SL-compatible clusters with
    $N_s=20,24,$ and 28; the upper and lower rows show $J_2=0$ and
    $J_2=0.25$, respectively. Blue and red disks denote positive and negative
    values of $\langle P_{j_{\rm ref},j}\rangle-1/4$, and their areas are
    proportional to the absolute value. The reference site is black. Boxed
    labels denote periodic images of sites in the principal cell, and green
    dashed arrows represent the torus vectors $\mathbf t_1$ and $\mathbf t_2$.}
    \label{fig:res:3}
\end{figure*}

The same conclusion is visible directly in the ground-state correlations. 
Figure~\ref{fig:res:3} displays the shifted correlations $\langle P_{j_{\rm ref},j}\rangle-\frac14$ (which gives $0$ in case there is no \SU{4} correlation between sites $j$ and $j_{\rm ref}$) on the three largest
4SL-compatible clusters. The alternating 4SL contrast (cf Fig. \ref{figJ2c} a) is present already at
$J_2=0$, but becomes substantially sharper at $J_2=0.25$. To quantify it, let
$\mathcal D_{4}$ be the six shortest directions connecting the reference site
to its own 4SL sublattice,
$\pm2\mathbf a_1$, $\pm2\mathbf a_2$, and
$\pm2(\mathbf a_2-\mathbf a_1)$, and let $\mathcal D_1$ be the six NN
directions. Then
\begin{equation}
\mathcal O_{4\mathrm{SL}}=
\frac{1}{6}\sum_{\mathbf d\in\mathcal D_4}
 \langle P_{j_{\rm ref},j(\mathbf d)}\rangle
-\frac{1}{6}\sum_{\mathbf d\in\mathcal D_1}
 \langle P_{j_{\rm ref},j(\mathbf d)}\rangle .
\label{eq:o4sl}
\end{equation}
Directions mapped onto the same site by periodic boundary conditions retain
their multiplicity. Figure~\ref{figJ2c}(d) shows that
$\mathcal O_{4\mathrm{SL}}$ remains finite over the available sizes and is
systematically enhanced by $J_2=0.25$. 

Taken together, the level crossings, the rotor branch and its vanishing slope,
the $N_s^{-3/2}$ energy scaling, and the real-space correlations consistently
support 4SL LRO at $J_2=0.25$ on clusters substantially larger than those of
Ref~\cite{mambrini2003}. The finite-size crossing data further suggest that a
nonzero NNN coupling is required, but the present sizes do not determine its
thermodynamic critical value.

Finally, in Tab.~\ref{tab:systems}, for future benchmarking, we give $12$ digits values of the minimal energies calculated for various clusters and symmetry sectors. 
In particular, as the extremal Ritz values in the Lanczos algorithm contain both the minimal and maximal energies, 
we were also able to check the mapping theorem (cf Section~\ref{theorem}) up to 12 digits for, e.g., 
the \SU{4} and \SU{7} singlet sector for the $N_s=28^{\Diamond\star}$ sites cluster (cf Tab.~\ref{tab:triangular_clusters}).

\begin{table}
\caption{\label{tab:systems}Irrep dimensions and associated ground-state energies for the nearest-neighbor Heisenberg model (cf also Tab.~\ref{tab:triangular_clusters} for the definitions of the clusters).}
\begin{ruledtabular}
\renewcommand{\arraystretch}{1.3}  % <-- increases all row heights
\begin{tabular}{llrl}
$N_s$ & irrep & dimension & $E_0$ \\
\hline
$21^{\star}$ & $[3, 3, 3, 3, 3, 3, 3]$ & 1'385'670   & -40.788693398762\\
$21^{\star}$ & $[4, 3, 3, 3, 3, 3, 2]$ & 19'953'648   & -38.562592108058\\
$24^{\Diamond}$ & $[6, 6, 6, 6]$ & 140'229'804   & -29.928066423294\\
$24^{\Diamond}$ & $[7, 6, 6, 5]$ & 1'262'068'236  & -28.106982447353\\
$28^{\Diamond\star}$ & $[7, 7, 7, 7]$ & 13'672'405'890 & -33.686668789411\\
$28^{\Diamond\star}$ & $[4, 4, 4, 4, 4, 4, 4]$ & 13'672'405'890 & -54.159024346190 \\
\end{tabular}
\end{ruledtabular}
\end{table}

\subsection{Long-range order for the pure \SU{7} HM}
\label{LRO7}

We now return to the nearest-neighbor model, $J_2=0$, and instead increase the
number of colors from four to seven. At first sight this route may appear
counterintuitive: more colors are often associated with stronger quantum
fluctuations and hence with a greater propensity for quantum-disordered
phases. The LFWT selection mechanism, however, depends on the local coloring
constraints rather than on this heuristic alone.

\begin{figure*}[t]
    \centering
    \includegraphics[width=\textwidth]{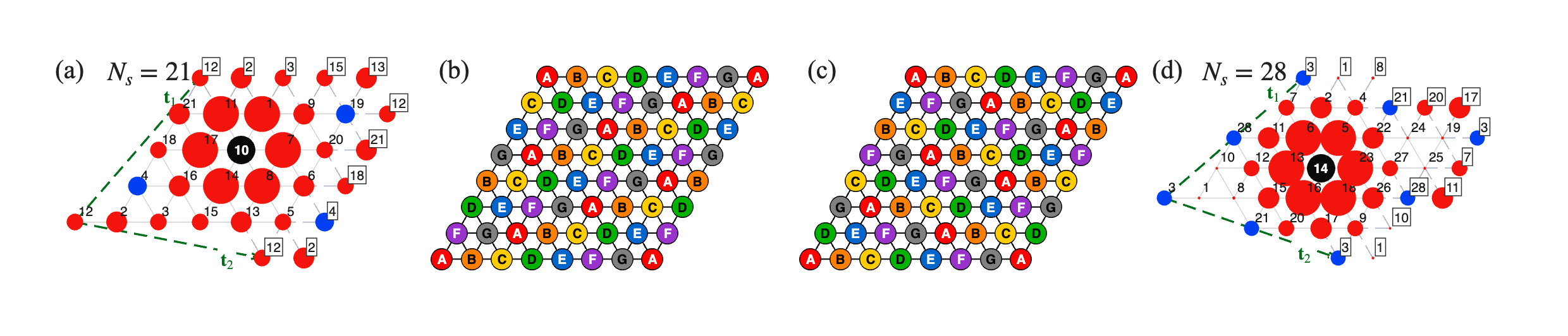}
    \caption{Seven-sublattice correlations and the two symmetry-related color
    patterns of the triangular-lattice \SU{7} HM.
    (a,d)~Ground-state permutation correlations
    $\langle P_{j_{\rm ref},j}\rangle-1/7$ on the compatible $N_s=21$ and
    $N_s=28$ clusters. Blue and red disks indicate positive and negative
    correlations, respectively, with areas proportional to their absolute
    values; the reference site is black. Boxed labels denote periodic images,
    and green dashed arrows show the torus vectors.
    (b,c)~The two mirror-related 7SL colorings. Their primitive translations
    are, respectively,
    $(-\mathbf a_1+3\mathbf a_2,\,3\mathbf a_1-2\mathbf a_2)$ and
    $(3\mathbf a_1-\mathbf a_2,\,-2\mathbf a_1+3\mathbf a_2)$.}
    \label{fig:corr_su7}
\end{figure*}

At the classical level, any site-factorized state in which NN sites carry
orthogonal colors has $\langle P_{i,j}\rangle=0$ on every bond and therefore
belongs to a macroscopically degenerate manifold. The harmonic zero-point
correction strongly reduces this degeneracy. At quadratic order, the energy is
minimized when a NN pair carrying two given colors is coupled only to sites
whose colors differ from both of them~\cite{Corboz_prx_2012}. On the TL, seven colors permit every site to have the six other colors as its
six neighbors. Up to a global permutation of the colors, the local condition
selects two periodic patterns\cite{LLM_contribution} related by a mirror reflection, shown in
Figs.~\ref{fig:corr_su7}(b) and \ref{fig:corr_su7}(c). Their primitive
translations can be chosen as
\begin{align}
(\mathbf T'_1,\mathbf T'_2)
 &=(-\mathbf a_1+3\mathbf a_2,\,3\mathbf a_1-2\mathbf a_2),\nonumber\\
(\mathbf T_1,\mathbf T_2)
 &=(3\mathbf a_1-\mathbf a_2,\,-2\mathbf a_1+3\mathbf a_2).
\label{eq:7sl_vectors}
\end{align}
These 7SL colorings are the TL counterparts of the \SU{5}
color configurations on the square lattice~\cite{nataf2014}. The stars in
Table~\ref{tab:triangular_clusters}
identify clusters whose torus translations are integer combinations of one of
the two pairs in Eq.~\eqref{eq:7sl_vectors}.

The correlations
$C_j^{(7)}=\langle P_{j_{\rm ref},j}\rangle-1/7$ on the compatible
$N_s=21$ and 28 clusters are displayed in Figs.~\ref{fig:corr_su7}(a) and~\ref{fig:corr_su7}(d). 
Their signs and spatial modulation reproduce the two
7SL patterns, providing a direct real-space signature of the order selected by
LFWT. 

\begin{figure}[ht]
    \centering
    \includegraphics[width=\linewidth]{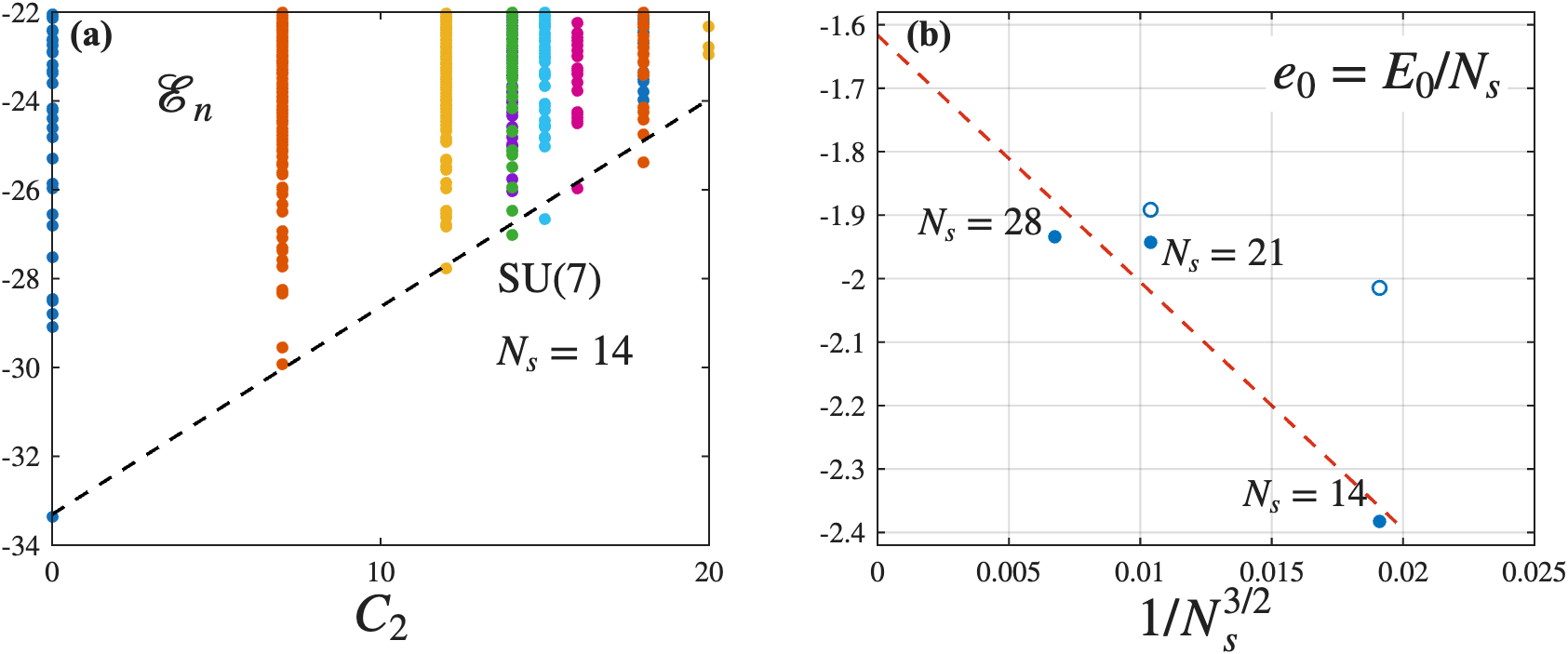}
    \caption{Finite-size signatures of seven-sublattice order in the
    nearest-neighbor \SU{7} HM.
    (a)~Low-energy spectrum of the $N_s=14$ cluster versus the quadratic
    \SU{7} Casimir. Colors distinguish the irreps, and the black dashed line
    is a fit to the lowest levels in the first three Casimir sectors.
    (b)~Ground-state energy per site versus $1/N_s^{3/2}$. Filled symbols are
    the $N_s=14,21,28$ clusters retained in the linear fit; the open
    $N_s=14$ and $N_s=21$ symbols denote incompatible clusters and are excluded.}
    \label{figATOSeoSU7}
\end{figure}

Two complementary spectral signatures are shown in
Fig.~\ref{figATOSeoSU7}. First, the $N_s=14$ low-energy levels organize into an
approximately linear tower when plotted against the quadratic \SU{7} Casimir.
Second, the ground-state energies of the three clusters retained in the fit
are well described by a linear dependence on $1/N_s^{3/2}$, while the open
14 and 21-site points, corresponding to incompatible clusters, lie away from this
trend. As in the \SU{4} case, this is the leading finite-size form expected
from linearly dispersing Goldstone modes in a two-dimensional N\'eel state.
The tower, the energy scaling, and the compatible-cluster correlations thus
form a coherent set of finite-size evidence for 7SL LRO in the pure \SU{7}
model. Because the fit contains only three sizes (i.e. $N_s=14, 21, 28$), larger clusters or
complementary methods remain necessary for a quantitative thermodynamic
extrapolation.

\section{Conclusion and outlook}

We have established a spectral mapping theorem for HMs with a
fundamental \SU{N} degree of freedom on each site. Transposition of SYTs maps
the singlet sector of \SU{N} onto that of \SU{N_s/N} and reverses the entire
energy level sequence within these sectors. For $N_s=28$, for example, the \SU{4}
singlet $[7,7,7,7]$ and the
\SU{7} singlet $[4,4,4,4,4,4,4]$ are transposed and have the same dimension,
$13\,672\,405\,890$. The mapping reverses the level ordering: at fixed
couplings, the antiferromagnetic ground state of one model corresponds to the
top of the spectrum of the other, or equivalently to the ground state after
reversing the signs of all exchanges within the singlet sector.

The bit-packed representation and efficient indexing of SYTs make this
theorem computationally useful and enable ED in symmetry sectors containing
more than ten billion states. On the TL, these developments do
not by themselves resolve the debated nearest-neighbor \SU{4} ground state.
They do, however, extend the evidence that NNN exchange stabilizes 4SL order
to much larger clusters than previously accessible. In the pure \SU{7} model,
the Anderson tower, ground-state-energy scaling, and real-space correlations
are all consistent with 7SL N\'eel order. This latter conclusion remains a
finite-size result and calls for independent simulations on larger systems.

Semistandard Young tableaux (SSYTs) provide symmetry-adapted bases for
\SU{N} Fermi--Hubbard Hilbert spaces in both
ED~\cite{Botzung_2023_PRL,Botzung_2023} and fully symmetry-adapted
DMRG~\cite{Nataf_2025_DMRG_FHM}. SYT formulations have also been used in
DMRG calculations for \SU{N} Heisenberg chains with fundamental and higher
local irreps~\cite{nataf_density_2018,gozel_2020,Nataf_2021}. These examples
motivate extending the present compact bit packing, ranking, and storage
schemes to SSYTs and tensor networks.

Finally, on the methodological side,
it would be interesting to take advantage of the spectral mapping to reinterpret the standard spins 1/2 antiferromagnetic HM on $N_s$ sites, 
as a ferromagnetic \SU{N_s/2} HM, allowing us to treat the thermodynamic limit of the spins 1/2 HM with large-$N$ techniques~\cite{Affleck_1985_largeN, Affleck_1988_largeN,Marston_1989_largeN}.

\begin{acknowledgments}
P.N. is supported by the CNRS Physique 2026 Tremplin grant.

Most of the computations presented in this paper
were performed using the GRICAD infrastructure
(\url{https://gricad.univ-grenoble-alpes.fr}),
which is supported by Grenoble research communities.

We also thank Profs. F. Mila and A. L\"auchli for useful advice.
\end{acknowledgments}

\section*{Data Availability Statement}
The code used in this study is available in Ref\cite{gozel_suned_2026}.

\appendix
\section{Computational implementation and performance}
\label{app:computational_details}

The bit-packed SYT representation introduced in Sec.~\ref{tech_aspec} is the basis of the distributed Lanczos solver used for the calculations reported in Sec.~\ref{LRO}. This appendix collects the implementation details, lattice-site ordering optimization, and performance benchmarks.

\subsection{Distributed-memory message passing}

We incorporate the bit-packed SYT representation into a custom Lanczos solver parallelized across compute nodes by message passing.
The ordered SYT basis is distributed among $N_{\mathrm{proc}}$ MPI processes, each of which constructs its own contiguous portion and therefore avoids communication during basis generation. In addition to its local basis elements, every process stores the smallest (or largest) SYT assigned to each process.
Given an SYT $Y^{[\nu]}_m$, a process can identify its owning rank $0\leqslant r<N_{\mathrm{proc}}$ by binary search in $\mathcal{O}(\log N_{\mathrm{proc}})$ operations. Together with Young's orthogonal rules, the indexing procedure, and standard Lanczos linear algebra, this ownership map provides the ingredients required for a distributed-memory ED implementation with full \SU{N} symmetry.

The memory saved by bit packing can instead be used to cache parts of the Hamiltonian. Storing the full Hamiltonian remains prohibitive, but its elementary building blocks, the adjacent tableau transpositions $\tau_{k,k+1}$, can be cached. For every such transposition, each process stores its list of off-diagonal SYT pairs and the corresponding axial distances $1/\rho$, thereby avoiding recomputation\footnote{An axial distance can be stored in a single byte.}. During a matrix-vector multiplication (MVM), each process sends the relevant Lanczos-vector element, multiplied by $\sqrt{1-\rho^2}$, to the process owning the image SYT.

\begin{figure}
    \centering
    \includegraphics[width=0.99\linewidth]{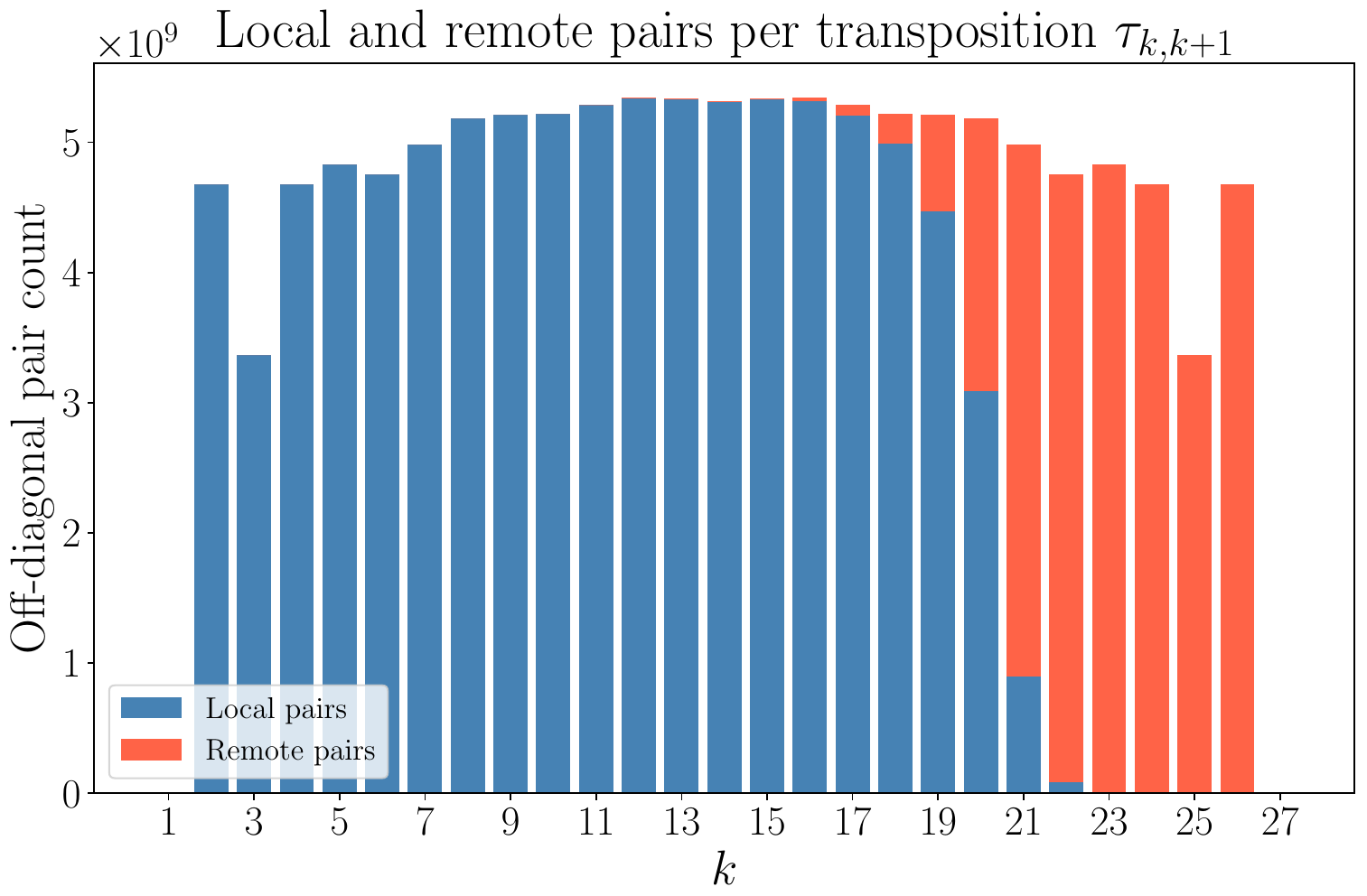}
    \caption{Distribution of off-diagonal SYT pairs between local pairs, whose two SYTs belong to the same MPI process, and remote pairs, whose SYTs belong to different processes. Results are shown for the $N_s=28$ singlet irrep of \SU{4}, distributed evenly over 192 processes. The transpositions $\tau_{1,2}$ and $\tau_{27,28}$ are diagonal.}
    \label{fig:pairs_counts}
\end{figure}

\begin{figure}
    \centering
    \begin{minipage}[t]{0.48\linewidth}
        \centering
        \includegraphics[width=\linewidth]{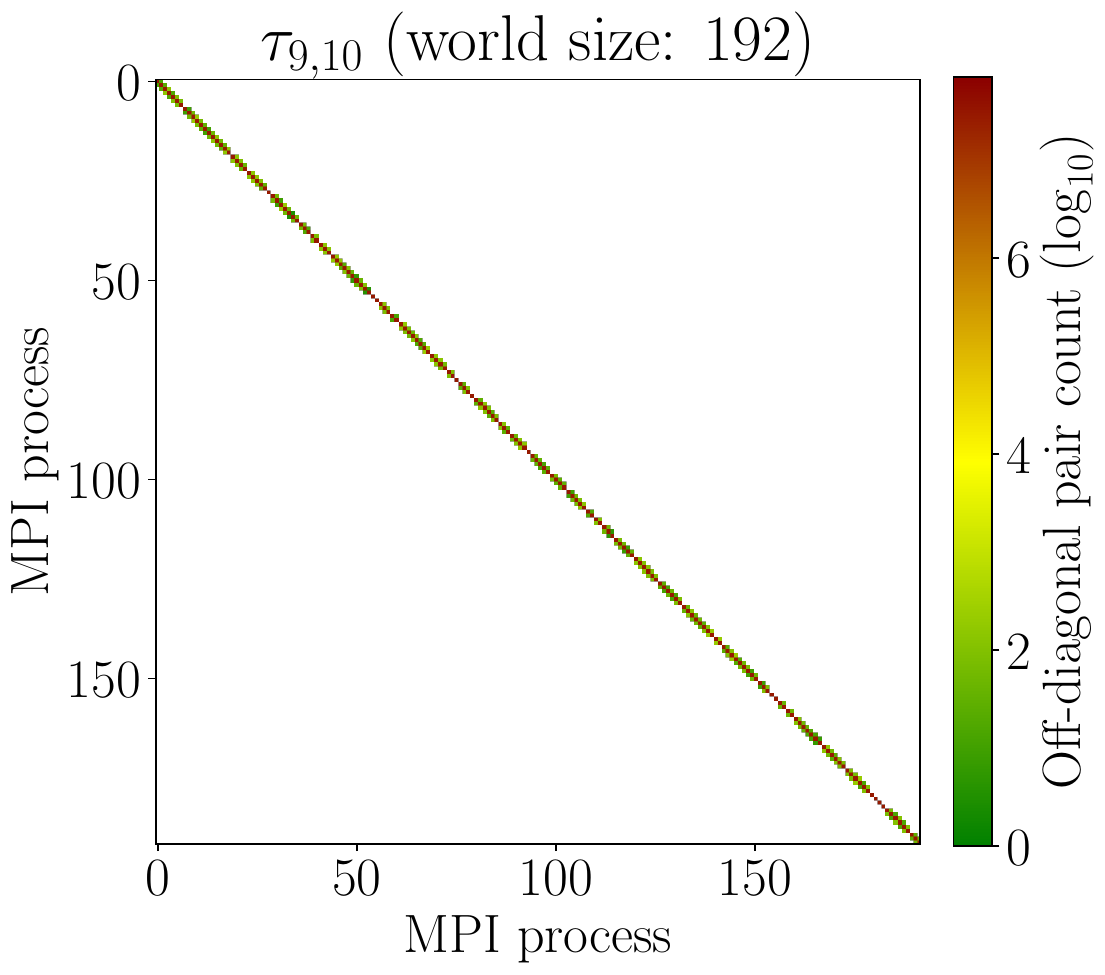}
    \end{minipage}
    \hfill
    \begin{minipage}[t]{0.48\linewidth}
        \centering
        \includegraphics[width=\linewidth]{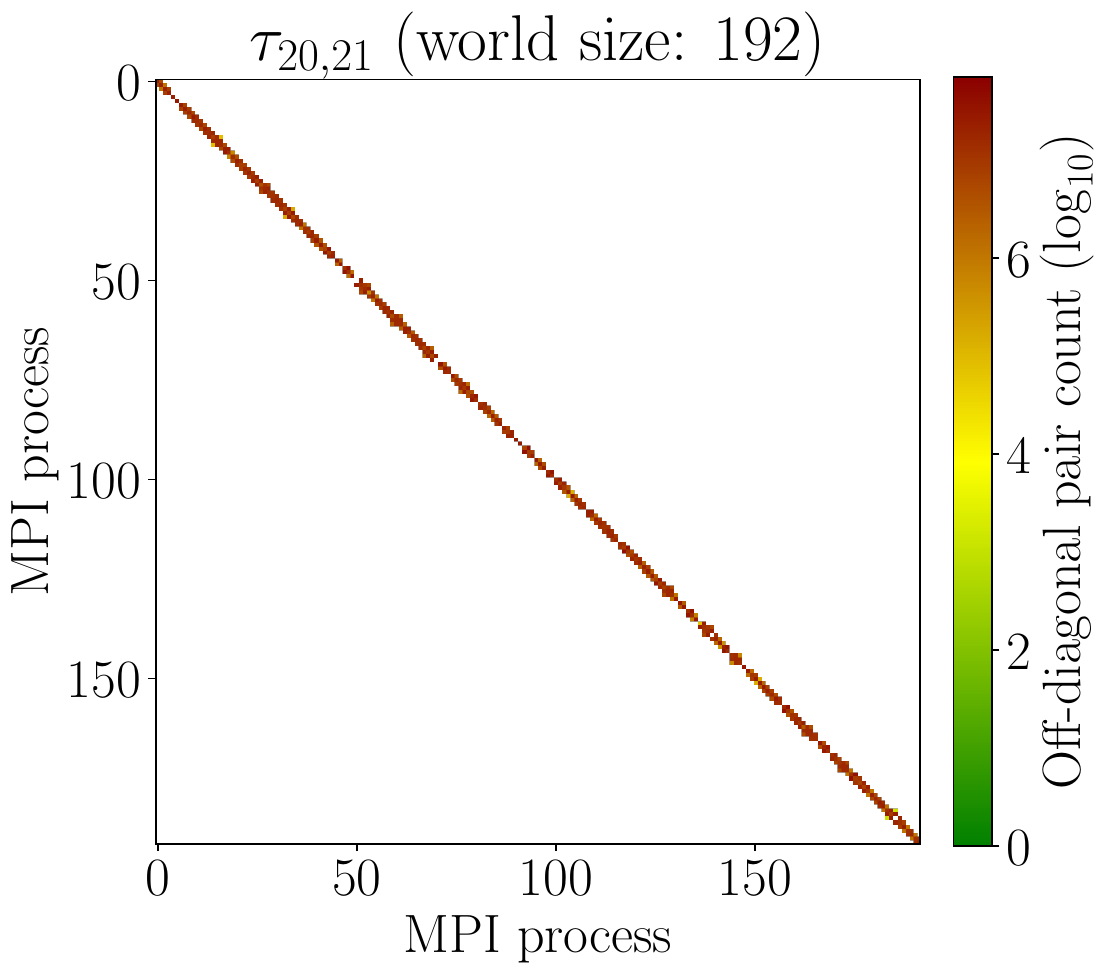}
    \end{minipage}

    \vspace{6pt}

    \begin{minipage}[t]{0.48\linewidth}
        \centering
        \includegraphics[width=\linewidth]{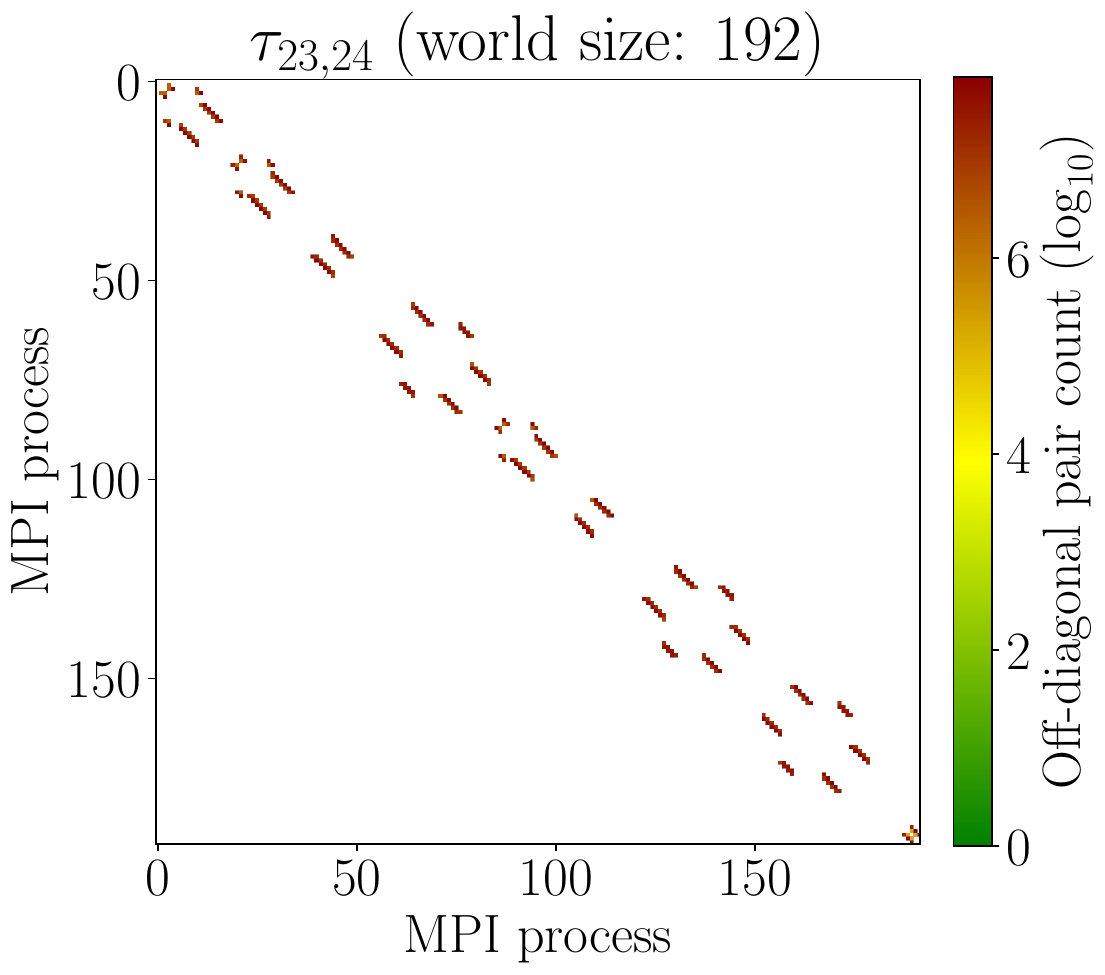}
    \end{minipage}
    \hfill
    \begin{minipage}[t]{0.48\linewidth}
        \centering
        \includegraphics[width=\linewidth]{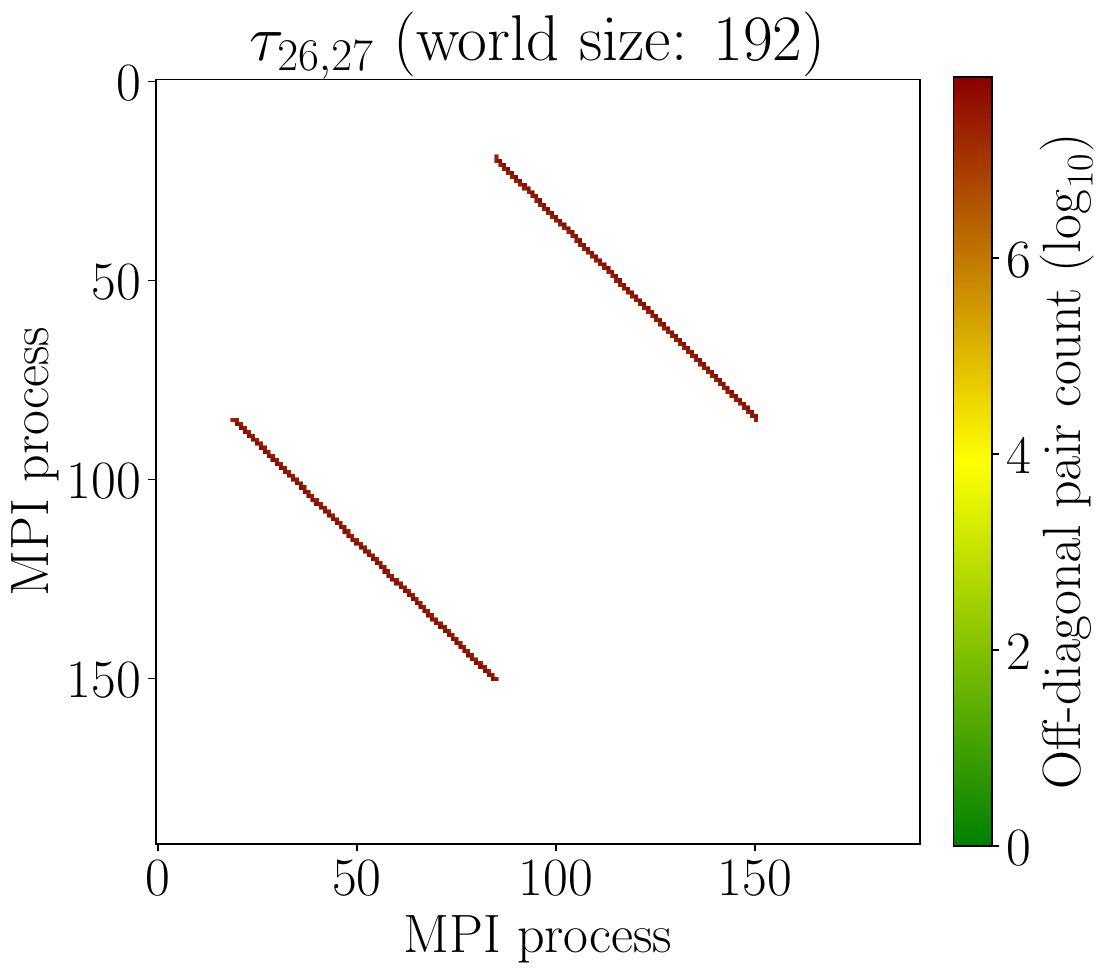}
    \end{minipage}

    \caption{Communication pattern and distribution of local and remote pairs for four representative transpositions. For $\tau_{9, 10}$ (top left), the vast majority of off-diagonal pairs are local to each process. For $\tau_{20, 21}$ (top right), remote pairs are almost as frequent as local pairs, but remote pairs belong to neighboring processes. For $\tau_{23, 24}$ and $\tau_{26, 27}$ (bottom left and bottom right, respectively), there is no local pair. Remote pairs are distributed among ``distant'' processes.}
    \label{fig:counts_k}
\end{figure}

The cost of applying $\tau_{k,k+1}$ is not uniform in $k$, because the number and distribution of its off-diagonal matrix elements vary. The transpositions $\tau_{1,2}$ and, for a singlet target irrep, $\tau_{N_s-1,N_s}$ are diagonal and therefore inexpensive. Every off-diagonal element pairs two SYTs through $\tau_{k,k+1}Y^{[\nu]}_m=Y^{[\nu]}_{m'}$. A pair is local when both SYTs belong to the same MPI process and remote otherwise; only remote pairs require an exchange of Lanczos-vector elements.
Figure~\ref{fig:pairs_counts} reports the numbers of local and remote pairs for every adjacent transposition in the $N_s=28$ singlet irrep of \SU{4}. Most transpositions contain predominantly local pairs, whereas the largest values of $k$ lead mainly to remote pairs. Figure~\ref{fig:counts_k} resolves the communication pattern for four representative transpositions. Even when all pairs are remote, communication involves only a small subset of processes rather than an all-to-all exchange. We therefore replace collective communication calls with dedicated asynchronous point-to-point operations.

\subsection{Optimization of the lattice-site ordering and strong-scaling performance}

The communication strategy reduces the cost of applying each adjacent transposition. 
A complementary optimization reduces the number of adjacent transpositions required to represent the Hamiltonian.

Each MVM in the Lanczos algorithm applies all $N_{\mathrm{bonds}}$ bond operators to a vector. A bond $P_{i,j}$ with $i<j$ is decomposed into $2|i-j|-1$ adjacent transpositions,
\[
P_{i,j}=P_{i,i+1}P_{i+1,i+2}\cdots P_{j-1,j}\cdots P_{i,i+1}.
\]
The total number of elementary operators is therefore $N_{\mathrm{ops}}=-N_{\mathrm{bonds}}+2\sum_b|i_b-j_b|$. 
Minimizing the MVM cost amounts to finding a labeling of the lattice sites that minimizes $N_{\mathrm{ops}}$, which is a linear-arrangement problem.
We developed a simulated-annealing code for this optimization~\cite{gozel_lattice_optimizer_2026}. 
For the periodic $N_s=28$ TL used here, a naive labeling leads to 1076 adjacent transpositions, whereas the optimized labeling, which appears in Fig. \ref{fig:res:3}, leads to 884. 
This 18\% reduction produces a comparable decrease in MVM runtime. 
Adding NNN interactions doubles the number of bonds but raises the optimized value of $N_{\mathrm{ops}}$ to 2368, a factor of $2.7$, with a corresponding increase in runtime. 
Table~\ref{tab:lattices} summarizes the clusters and Hamiltonians used in the benchmarks.

\begin{table}
\caption{\label{tab:lattices}Triangular lattices with $N_s = 21, 24, 28$ sites and underlying Hamiltonians.}
\begin{ruledtabular}
\renewcommand{\arraystretch}{1.3}  % <-- increases all row heights
\begin{tabular}{llll}
$N_s$ & Hamiltonian & $N_{\mathrm{bonds}}$ & $N_{\mathrm{ops}}$ \\
\hline
$21^{\star}$ & Pure Heisenberg & 63 & 561 \\
$24^{\diamond}$ & Pure Heisenberg & 72 & 672 \\
$24^{\diamond}$ & Heisenberg + NNN & 144 & 1784 \\
$28^{\diamond\star}$ & Pure Heisenberg & 84 & 884 \\
$28^{\diamond\star}$  & Heisenberg + NNN & 168 & 2368 \\
\end{tabular}
\end{ruledtabular}
\end{table}

Having reduced both the communication overhead per transposition and the number of transpositions per MVM, we now assess the resulting strong-scaling performance.

\begin{figure}
    \centering
    \includegraphics[width=0.99\linewidth]{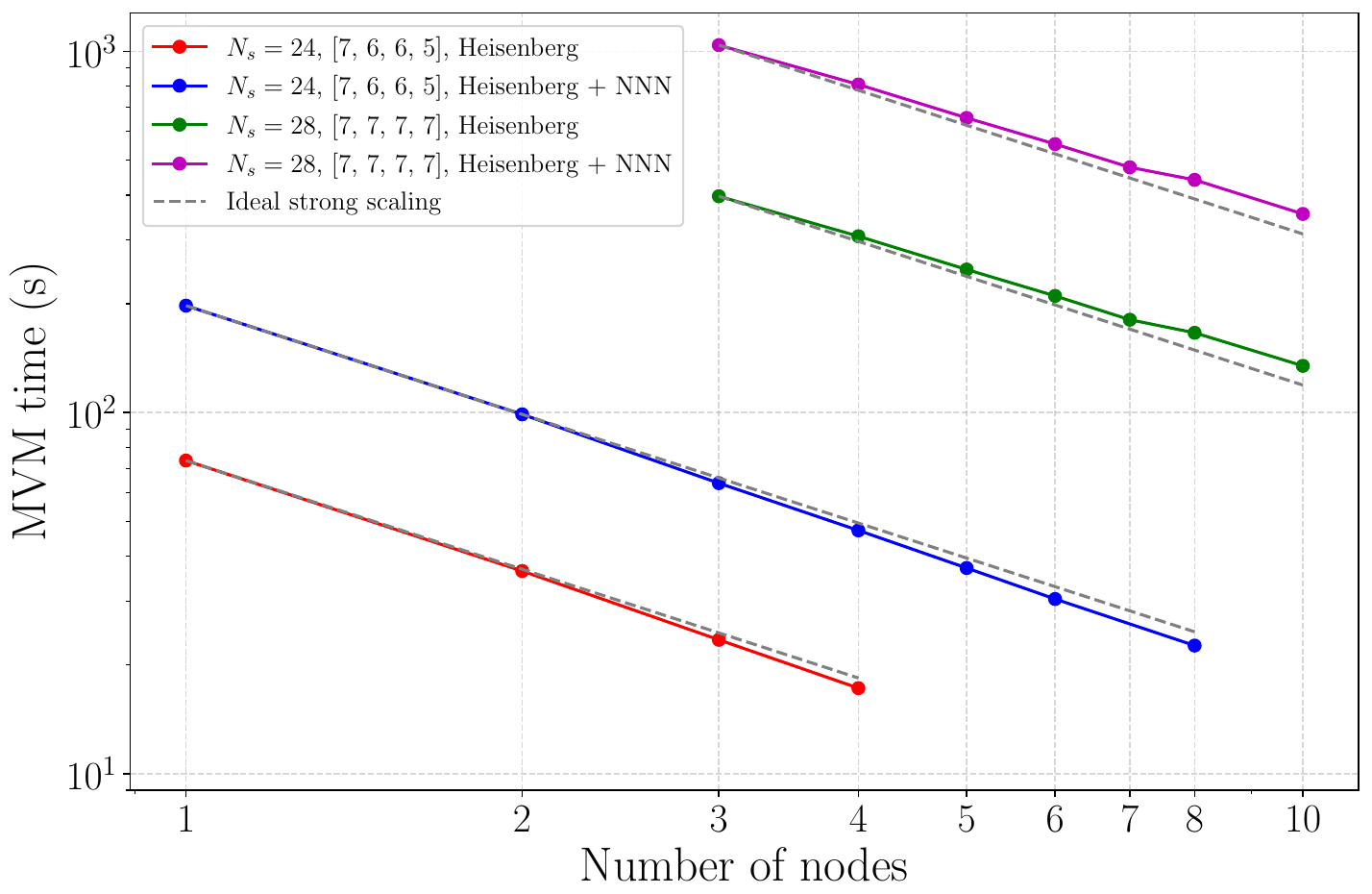}
    \caption{Strong scaling of the matrix-vector multiplication (MVM) runtime with the number of compute nodes for the $N_s=24$ HM in the adjoint irrep of dimension $1.26$~billion and the $N_s=28$ model in the singlet irrep of dimension $13.67$~billion, with and without next-nearest-neighbor (NNN) interactions. Each node contains two AMD EPYC 9654 96-core processors, for a total of 192 cores per node.}
    \label{fig:mvm_strong_scaling}
\end{figure}

Figure~\ref{fig:mvm_strong_scaling} shows the strong scaling of the MVM runtime with the number of compute nodes for the four representative systems summarized in Table~\ref{tab:systems}.

\bibliographystyle{apsrev4-1}
\bibliography{bibliography_triangle}

\end{document}